\documentclass[fleqn,usenatbib]{mnras}

\usepackage[T1]{fontenc}
\usepackage{ae,aecompl}

\usepackage[UKenglish]{babel}
\usepackage[babel=true]{csquotes}  

\usepackage[final]{microtype}
\usepackage{newtxtext} 
\usepackage{amsmath}	
\usepackage{newtxmath} 
\usepackage{amssymb}	
\usepackage{upgreek}	
\usepackage{ulem}       

\usepackage{graphicx}	

\usepackage{subfig}
\usepackage{booktabs}	
\usepackage{multirow}	

\usepackage[capitalise, nameinlink, noabbrev]{cleveref}
\creflabelformat{equation}{#2#1#3}	

\hypersetup{pdfencoding=unicode,}

\newcommand{\img}[1]{./img/#1.pdf}

\defcitealias{KRUI14}{KL14}
\newcommand{\KLPrin}{\citetalias{KRUI14} principle}

\newcommand*{\VerticalCentered}[1]{%
	\begingroup%
	\setbox0=\hbox{#1}%
	\parbox{\wd0}{\box0}%
	\endgroup%
}

\newcommand*{\TableFootnote}[2]{%
	\begingroup%
	\sbox0{#1}
	\mbox{\parbox{\dimexpr\wd0 - 10pt \relax}{#1\\#2}}
	\endgroup%
}

\newcommand{\Ha}{H$\alpha$}

\usepackage{etoolbox}

\title[Star formation rate tracer lifetimes]{An uncertainty principle for star formation -- III.\ The characteristic emission time-scales of star formation rate tracers}

\author[Daniel~T.~Haydon et al.]{%
	Daniel~T.~Haydon,\textsuperscript{1}
	J.~M.~Diederik Kruijssen,\textsuperscript{1,2}\thanks{E-mail: kruijssen@uni-heidelberg.de}
	M\'{e}lanie~Chevance,\textsuperscript{1}
	\vspace{-1mm}Alexander~P.~S.\newauthor
	Hygate,\textsuperscript{2,1}
	Mark~R.~Krumholz,\textsuperscript{3,4}
	Andreas~Schruba,\textsuperscript{5}
	and Steven~N.~Longmore\textsuperscript{6}
	\\
	\vspace{-1mm}\textsuperscript{1}Astronomisches Rechen-Institut, Zentrum f{\"u}r Astronomie der Universit{\"a}t Heidelberg, M{\"o}nchhofstra{\ss}e 12--14, 69120 Heidelberg, Germany\\
	\vspace{-1mm}\textsuperscript{2}Max-Planck-Institut f{\"u}r Astronomie, K{\"o}nigstuhl 17, 69117, Heidelberg, Germany\\
	\vspace{-1mm}\textsuperscript{3}Research School of Astronomy and Astrophysics, Australian National University, Canberra 2611, A.C.T., Australia\\
	\vspace{-1mm}\textsuperscript{4}Centre of Excellence for Astronomy in Three Dimensions (ASTRO-3D), Australia\\
	\vspace{-1mm}\textsuperscript{5}Max-Planck Institut f{\"u}r Extraterrestrische Physik, Giessenbachstra{\ss}e 1, 85748 Garching, Germany\\
	\vspace{-1mm}\textsuperscript{6}Astrophysics Research Institute, Liverpool John Moores University, IC2, Liverpool Science Park, 146 Brownlow Hill, Liverpool L3 5RF, United Kingdom
}

\date{Accepted 2020 August 11. Received 2020 July 23; in original form 2018 September 14}

\pubyear{2020}

\begin{document}
	\label{firstpage}
	\pagerange{\pageref{firstpage}--\pageref{lastpage}}
	\maketitle

	\begin{abstract}
		We recently presented a new statistical method to constrain the physics of star formation and feedback on the cloud scale by reconstructing the underlying evolutionary timeline.
		However, by itself this new method only recovers the \textit{relative} durations of different evolutionary phases.
		To enable observational applications, it therefore requires knowledge of an absolute \enquote{reference time-scale} to convert relative time-scales into absolute values.
		The logical choice for this reference time-scale is the duration over which the star formation rate (SFR) tracer is visible because it can be characterised using stellar population synthesis (SPS) models.
		In this paper, we calibrate this reference time-scale using synthetic emission maps of several SFR tracers, generated by combining the output from a hydrodynamical disc galaxy simulation with the SPS model \textsc{slug2}.
        We apply our statistical method to obtain self-consistent measurements of each tracer's reference time-scale.
        These include H${\alpha}$ and 12 ultraviolet (UV) filters (from GALEX, Swift, and HST), which cover a wavelength range 150--350~nm.
		At solar metallicity, the measured reference time-scales of H${\alpha}$ are ${4.32^{+0.09}_{-0.23}}$~Myr with continuum subtraction, and 6--16~Myr without, where the time-scale increases with filter width.
        For the UV filters we find 17--33~Myr, nearly monotonically increasing with wavelength.
		The characteristic time-scale decreases towards higher metallicities, as well as to lower star formation rate surface densities, owing to stellar initial mass function sampling effects.
		We provide fitting functions for the reference time-scale as a function of metallicity, filter width, or wavelength, to enable observational applications of our statistical method across a wide variety of galaxies.
	\end{abstract}



	\section{Introduction}
	It is a challenging problem in astrophysics to characterise the time-scales over which astrophysical processes take place, because most of these processes take much longer than a human lifetime.
	For this reason, star formation studies have struggled to identify the physical processes governing the evolution of molecular clouds and star forming regions, which requires knowledge of the underlying time-scales \citep[e.g.][]{DOBB14, KRUM14,CHEV20c}.
    Traditionally, the age of the stellar population has been used as a \enquote{reference time-scale} to infer how long other phases of the star formation lifecycle take, such as the molecular cloud lifetime and the time over which gas and young stars are associated \citep[e.g.][]{LEIS89, ELME00, KAWA09}.
    However, the time-scales measured to date span a wide range of durations \citep[e.g.][]{SCOV79, KODA09, MEID15}, which is largely caused by the heterogeneity of the methods used \citep[see the discussion in the Methods section of][]{KRUI19}.
    Clearly, a systematic framework is needed for placing the lifecycle of molecular clouds, star formation, and feedback on an absolute, empirically-determined evolutionary timeline.

	In \citet{KRUI14}, we put forward a new statistical method, titled \enquote{an uncertainty principle for star formation} (hereafter \KLPrin{}{}), formalised in the the \textsc{Heisenberg} code \citep{KRUI18}. This analysis method enables the use of pairs of high-resolution emission maps tracing successive phases of the evolutionary cycle between cloud evolution, star formation, and feedback (e.g.\ CO tracing molecular gas and H${\alpha}$ tracing recent star formation) to infer the relative durations of these phases on the cloud scale.
    Specifically, this method can be used to measure the relative duration of the \enquote{cloud lifetime} and the relative duration over which a molecular cloud is disrupted by stellar feedback, compared to the characteristic \enquote{H\,\textsc{ii} region lifetime}.

	To turn these resulting \textit{relative} durations into an absolute timeline, it is critical to normalise the timeline using a known \enquote{reference time-scale}.
    Without this reference time-scale, observational applications of the \KLPrin{} cannot be used to obtain meaningful constraints on the cloud lifecycle.
	The most direct way of providing a reference time-scale is by following the example of previous work in this area and characterising the time-scales of star formation rate (SFR) tracers.
	Indeed, our new methodology is preceded (and partially inspired) by a wide variety of literature aiming to observationally characterise the cloud lifecycle, which all used the approximate lifetimes of H\,\textsc{ii} regions or the ages of young stellar clusters as a reference time-scale \citep[e.g.][]{BLIT07,KAWA09,MIUR12,CORB17}.
	In this paper, we calibrate this approach for use in observational applications of the \KLPrin{}.
	Doing so requires a controlled experiment, in which the duration of one phase is known exactly and the duration of the other phase is measured from its emission map using \textsc{Heisenberg}.
    This requires the use of galaxy simulations rather than real galaxies.
	Since the reference time-scale likely depends on metallicity and is affected by the sampling of the stellar initial mass function (IMF), the experiment must also be repeated for different tracers, such as H${\alpha}$ and various ultraviolet (UV) filters, metallicities, and degrees of IMF sampling.

	Even though it is the goal of this paper to define the tracer time-scales within the context of the \KLPrin{}, measuring the characteristic emission time-scales of SFR tracers is also important in other contexts.
	For instance, this characteristic time-scale is an indicator of the duration for which photoionising feedback can act on the surrounding interstellar medium.
	Deriving absolute SFRs from observed line or broadband emission flux also requires calibration by the time-scale over which a young stellar population emits at that given wavelength.
	The emission in different wavebands is dominated by different types of stars \citep{HAO11} and so it is possible to determine relationships between the luminosity at a given wavelength and the SFR based on the lifetimes of these stars \citep[e.g.][]{CALZ07, HAO11, MURP11, KENN12}.
	For example, producing H${\alpha}$ emission requires ionising photons from very massive stars, which have lifetimes ${<10}$~Myr \citep{LEIT99, MURP11}, such that the emission itself should fade on a characteristic time-scale that is of a similar magnitude.

	We emphasise that by describing the duration of SFR tracer emission with a single time-scale, we do not assume that the SFR tracer emission sharply drops at some particular age.
	Instead, we describe the gradual fading of emission in terms of a single time-scale that is meaningful in the context of our statistical method.
	Conceptually, this can be regarded as analogous (but not equal) to the e-folding time of an exponential decay, or the time at which 50~per~cent of the total H${\alpha}$ luminosity that will be produced by a young stellar population has been emitted.
	This defines a single number for the time-scale of emission for that population.
	Since the emission fades gradually, it depends on the physical context what definition of a single time-scale is correct.
	Therefore, our goal is not to provide a general definition of the time-scale of SFR tracer emission, but to provide the right definition of an SFR tracer time-scale for use in observational applications of the \KLPrin{}.

	Previous work attempting to derive characteristic time-scales for different SFR tracers has revealed that the major problem obstructing a conclusive measurement is that there is no obvious definition of the time-scale that should be adopted.
	Instead, there exists a range of possible definitions, such as a luminosity-weighted mean, a percentage intensity change, or a percentage of the cumulative emission.
    The choice of definition can result in differences of up to an order of magnitude in time-scale \citep{LERO12, KENN12}.
	In view of this strong dependence on the precise definition of the reference time-scale, we opt to use a self-consistent approach for determining the SFR tracer time-scale; that is, we measure the emission time-scales of SFR tracers by applying the \KLPrin{} itself to synthetic emission maps, which have been generated by combining the output from a hydrodynamical disc galaxy simulation with the stellar population synthesis (SPS) model \textsc{slug2} \citep{SILV12, SILV14, KRUM15}.
	The reference time-scales obtained this way critically enable observational applications of the \KLPrin{}, which provides measurements of a wide variety of physical quantities as part of a single analysis \citep{KRUI18}, such as the molecular cloud lifetime, the time-scale for cloud destruction by feedback, the separation length between independent star-forming regions, the integrated cloud-scale star formation efficiency, and the feedback outflow velocity.
	The method has also been extended to also provide physically-motivated measurements of the diffuse gas fraction \citep[e.g.\ of molecular, atomic, or ionised gas;][]{HYGA19}.
	For the first applications of the method measuring these quantities across nearby star-forming galaxies, we refer the reader to \citet{KRUI19} and \citet{CHEV20}.

    The structure of this paper is as follows.
	In \cref{sec:KL14}, we summarise the \KLPrin{} and the practical application of the associated \textsc{Heisenberg} code.
	We outline our approach for constraining the characteristic time-scales of different SFR tracers with well-sampled IMFs in \cref{sec:methodNonStoc}.
    For solar metallicity, the resulting time-scales are presented in \cref{sec:resultsSingleTimeScale}.
	In \cref{sec:Metallicity}, we demonstrate how the time-scales depend on metallicity.
	In \cref{sec:IMFsampling}, we demonstrate the effects of incomplete IMF sampling, which is expected to change the results in environments of low SFR surface density.
	In \cref{sec:obs}, we carry out a brief test of the obtained time-scales, by comparing them to observations of H${\alpha}$ and UV emission in NGC300.
	Finally, we summarise the results and present our conclusions in \cref{sec:conclusions}.

	\section{Uncertainty principle for star formation}\label{sec:KL14}
	The analysis presented in this paper is based on the \KLPrin{} and its specific realisation in the \textsc{Heisenberg} code \citep{KRUI18}.
	The statistical method presented by \citet{KRUI14} and \citet{KRUI18} enables the characterisation of the cloud-scale physics of star formation and feedback in a systematic way, based on the spatial distribution of emission in pairs of maps tracing particular evolutionary stages of the star formation process.
	We briefly describe the concept of the method and summarise its specific application in this work, aimed at constraining the characteristic time-scales of SFR tracers.

	The \KLPrin{} has been primarily used to determine the lifetime of giant molecular clouds in nearby galaxies, by comparing the spatial distributions of the emission of a molecular gas tracer (e.g.\ CO) and SFR tracer (e.g.\ \Ha ).
	The first applications of this method \citep{KRUI19,HYGA19b,CHEV20b,CHEV20, WARD20b, ZABE20} demonstrate that clouds are highly dynamic and, once they host unembedded massive stars, are dispersed on time-scales shorter than the typical supernova delay time, implying that early feedback by photoionisation or stellar winds plays a critical role in disrupting molecular clouds.
	However, the applicability of the method is not restricted to maps of molecular gas or young stellar emission.
	Depending on the combination of tracers used, it is possible to constrain the durations of different stages of the star formation timeline using \textsc{Heisenberg}, such as the atomic cloud lifetime \citep{WARD20} or the duration of the embedded phase of star formation \citep{KIM20}.
	
    In basic terms, the \KLPrin{} represents a galaxy as a collection of independent (e.g.\ star-forming) regions, where each region is evolving along its timeline independently of its neighbouring regions.
	The number of regions that are emitting in each of the two tracers (with the possibility that some regions are in a transition phase and emit in both tracers) is roughly related to the duration of that phase -- the shorter the duration of a phase, the less likely it is to observe a region in that phase.

	Most importantly in the context of this paper, this method fundamentally derives the duration of one phase \textit{relative} to another one. The duration of one of the two phases must therefore be known a priori in order to derive \textit{absolute} time-scales.
    In practice, this \enquote{reference time-scale} can often be associated with the duration of the emission of the SFR tracer, owing to the absolute clock provided by stellar evolution \citep[e.g.][]{LEIT99, LEIT14}.
    In this paper, we use this clock to measure the reference time-scale of SFR tracers and thereby provide a calibration of the evolutionary timeline between molecular clouds, star formation and feedback that can be measured observationally using \textsc{Heisenberg}.
	Throughout this work, we call \enquote{reference map} the emission map associated to the phase of known duration, and we refer to this duration as the \enquote{reference time-scale}.

	\begin{figure}
		\centering
		\includegraphics[width=\columnwidth]{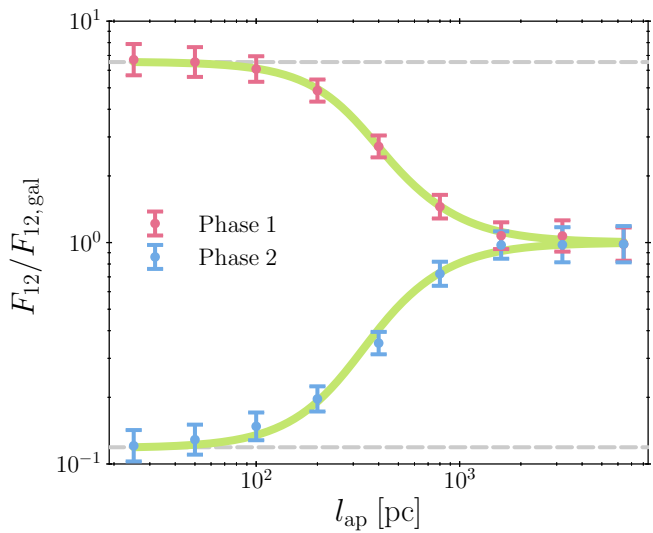}%
		\caption{%
			Example tuning fork diagram produced by the \textsc{Heisenberg} code.
			The figure shows the relative change of the phase-1-to-phase-2 flux ratio (${F_{12}}$) when placing apertures of size ${l_{\mathrm{ap}}}$ at the locations of emission peaks in the phase 1 map (top branch, in red) or the phase 2 map (bottom branch, in blue), compared to the galactic-scale phase-1-to-phase-2 flux ratio (${F_{\mathrm{12,\,gal}}}$) as a function ${l_{\mathrm{ap}}}$.
			The best-fitting model is indicated by the two green curves.
		}\label{fig:exampleTuningfork}
	\end{figure}
	
	We briefly summarise the procedure used by \textsc{Heisenberg} and refer the reader to \citet{KRUI18} for the specific details.
	The method relies on the findings of \citet{KRUI14}, where it is shown that the gas-to-young stellar flux ratio changes relative to the galactic average when focusing apertures on peaks of gas or young stellar emission, and that this relative change is a direct function of the underlying evolutionary timeline describing how gas is converted into stars on the cloud scale.
	The procedure is as follows.

	\begin{enumerate}
		\item We use two emission maps of the same galaxy, tracing two successive evolutionary phases (hereafter phase 1 and phase 2).
        The phase 2 map is used as the reference map, with its duration equal to the reference time-scale, ${t_{\mathrm{R}}}$. 
        In this work, we construct a map of star particles within a given age range as the phase 2 map of known duration (see \cref{sec:methodNonStoc}) and provide a synthetic SFR tracer map as phase 1, of which we measure the duration (see below).	
		\item Each map is convolved using a top hat kernel for a series of aperture sizes (ranging from $\sim$ the cloud scale to $\sim$ the galaxy scale).
		\item For each of these convolution scales, apertures of corresponding sizes are placed on the emission peaks of both convolved maps. The enclosed phase-1-to-phase-2 flux ratio (${F_{12}}$) within these apertures is then measured in units of the galactic average flux ratio (${F_{\mathrm{12,\,gal}}}$). 
		On the small scales, the measured flux ratio in apertures centred on phase 1 peaks (respectively phase 2 peaks) deviates from the galactic average value, as visible in \cref{fig:exampleTuningfork}.
		\item The models describing the shapes of the two branches of this \enquote{tuning fork diagram} \citep[see][Equations~81 and 82]{KRUI18} are fitted to the measurements in order to constrain the following three free parameters: the typical separation length between identified peaks (${\lambda}$), the relative temporal overlap between the two phases (${t_{\mathrm{over}}/t_{\mathrm{R}}}$), and the relative duration of phase 1 (${t_{\mathrm{1}}/t_{\mathrm{R}}}$).
		\footnote{%
			These quantities represent flux-weighted averages across the population of emission peaks (see Section~3.2.9 and~3.2.11 of \citealt{KRUI18} and Equation~1 of \citealt{KRUI19}).
			In addition, ${\lambda}$ describes the typical separation length in the vicinity of peaks, not the area-averaged value across an entire galaxy, making it relatively insensitive to morphological features such as spiral arms \citep{KRUI18,KRUI19,CHEV20}.
		}
		The absolute time-scales ${t_{\mathrm{1}}}$ and ${t_{\mathrm{over}}}$ require a reference time-scale ${t_{\mathrm{R}}}$.
	\end{enumerate}
	
    The reference time-scale is a key ingredient for retrieving the absolute time-scales ${t_{\mathrm{1}}}$ and ${t_{\mathrm{over}}}$. 
	In this paper, we use the above method to calibrate the duration of emission for a variety of SFR tracers, which can then be used as the reference time-scale in observational applications.
	In \cref{sec:methodNonStoc}, we describe how we use a simulated galaxy to create reference maps from stars within a known age bin (the duration of which is used as ${t_{\mathrm{R}}}$), and synthetic SFR tracer emission maps.
	By adopting one of the synthetic SFR tracer emission maps as the phase 1 map and a reference map with a known age interval as the phase 2 map, we have (by definition) two maps which trace successive phases of an evolutionary timeline. We then use \textsc{Heisenberg} to constrain their relative lifetimes as well as the absolute duration of phase 1, ${t_{\mathrm{1}}}$.

	The above approach implicitly assumes that the method is commutative and transitive, i.e.\ we can use stars in a known age range (A) to calibrate an SFR tracer time-scale (B), which then is used in observational applications to characterise e.g.\ the molecular cloud lifetime (C), so that effectively the known age range (A) is used to calibrate the cloud lifetime (C). Both the commutativity \citep{KRUI18} and transitivity \citep{WARD20} of the method have been demonstrated in other papers, which justifies its use in this work. The fact that the method is transitive and commutative follows somewhat trivially from its ability to predict correct lifetimes, as demonstrated by \citet{KRUI18}. If the method correctly predicts a time-scale ratio $t_{\rm A}/t_{\rm B}$, then it must also correctly predict the time-scale ratio $t_{\rm B}/t_{\rm A}$. Likewise, if the method correctly predicts the time-scale ratios $t_{\rm A}/t_{\rm B}$ and $t_{\rm B}/t_{\rm C}$, then it must also correctly predict the time-scale ratio $t_{\rm A}/t_{\rm C}$.

	\section{Method for calculating the characteristic emission time-scales of SFR tracers for a fully-sampled IMF}\label{sec:methodNonStoc}
	\begin{table}
		\caption{%
			We derive characteristic time-scales for the SFR tracers listed here.
		}\label{tab:filterList}
		\subfloat[%
            The UV filters we consider.
            ${\overline{\lambda}_{\mathrm{w}}}$ is the response-weighted mean wavelength of the filter.
            The normalised filter response curves are presented in \cref{sec:resultsSingleTimeScale}.\label{tab:UVfilterList}
        ]{%
			\begin{minipage}{\columnwidth}
				\centering
				\begin{tabular}{lllc}
					\toprule
					Telescope & Instrument & Filter & ${\overline{\lambda}_{\mathrm{w}}}$ ${\left[\mathrm{nm}\right]}$\\
					\midrule
					GALEX & & FUV & 153.9\tabularnewline
					GALEX & & NUV & 231.6\tabularnewline
					Swift & UVOT & M2 & 225.6\tabularnewline
					Swift & UVOT & W1 & 261.7\tabularnewline
					Swift & UVOT & W2 & 208.4\tabularnewline
					HST & WFC3 & UVIS1 F218W & 223.3\tabularnewline
					HST & WFC3 & UVIS1 F225W & 238.0\tabularnewline
					HST & WFC3 & UVIS1 F275W & 271.5\tabularnewline
					HST & WFC3 & UVIS1 F336W & 335.8\tabularnewline
					HST & WFPC2 & F255W & 259.5\tabularnewline
					HST & WFPC2 & F300W & 297.4\tabularnewline
					HST & WFPC2 & F336W & 335.0\tabularnewline
					\bottomrule
				\end{tabular}
			\end{minipage}
		}\\%
		\subfloat[The H${\alpha}$ filters we consider.\label{tab:HAfilterList}]{%
			\begin{minipage}{\columnwidth}
				\centering
				\begin{tabular}{p{0.15\columnwidth} p{0.75\columnwidth}}
					\toprule
					Filter & Details\\
					\midrule
					H${\alpha{-}}$ & H${\alpha}$ emission with continuum subtraction.
                    This is not a true filter but a direct measurement of the hydrogen-ionizing photon emission, see \cref{sec:emissionMap} for details.\\
					${\mathrm{H\alpha{+}}\,W}$ & A narrow band filter including H${\alpha}$ and the continuum as defined in \cref{eq:HAFilter}.
					The total filter width is indicated by ${W}$; we consider ${W = \left\{10,~20,~40,~80,~160\right\}~\text{\AA{}}}$.\\
					\bottomrule
				\end{tabular}
			\end{minipage}
		}
	\end{table}

	We present here the steps we take to find the characteristic time-scales for H${{\alpha}}$ and UV SFR tracers (see \cref{tab:filterList} for details) using synthetic emission maps and the \textsc{Heisenberg} code.
	As we described in \cref{sec:KL14}, \textsc{Heisenberg} can determine the duration of the first input map from the second by using the latter as a reference map (i.e.\ the map showing the evolutionary phase of known duration).
	This means that if we provide \textsc{Heisenberg} with a galaxy map of one of the SFR tracers (e.g.\ H${{\alpha}}$) along with a reference map, \textsc{Heisenberg} can provide us with the time-scale associated with that SFR tracer.
	This approach to measure the SFR tracer time-scales ensures that the obtained reference time-scales are self-consistent within the context of our method.
	After all, the SFR tracer will be applied as the reference time-scale in future observational applications of \textsc{Heisenberg}.

	We generate both the SFR tracer maps and the reference maps using a numerical simulation of a flocculent spiral galaxy.
    Fundamentally, we only require some (preferably physically-motivated) correlation of positions and ages of star particles to carry out the experiments of this paper, implying that we could have used any (e.g.\ randomly-generated) distribution of points or Gaussian-like regions.
    However, the use of a galaxy simulation is more physically appropriate, as it contains some imprint of galactic morphology and the positional correlation of star formation events as a result of self-gravity and stellar feedback.
	Using a galaxy simulation still carries the advantage that we have complete control over the duration of the reference map, by using stellar particles of a specified age range.
	The SFR tracer maps are generated using a SPS model.
	This approach allows us to additionally quantify the effects of metallicity (see \cref{sec:Metallicity}) and IMF sampling (see \cref{sec:IMFsampling}) on the SFR tracer time-scale.
	In turn, this will facilitate observational applications of \textsc{Heisenberg} to a variety of galactic environments.

	We discuss the adopted galaxy simulation in \cref{sec:simulation}, the method for generating the reference maps in \cref{sec:refMap}, and the method for generating the synthetic SFR tracer maps in \cref{sec:emissionMap}.

	\subsection{Galaxy simulation}\label{sec:simulation}
	The results in this paper are based on the \enquote{high-resolution} simulated galaxy from \citet{KRUI18}.
	We set up the initial conditions for this galaxy using the methods described in \citet{SPRI05}.
	The simulation has a total of ${{4.95\times10^{6}}}$ particles: ${1\times10^{6}}$ in the dark matter halo, ${2.31\times10^{6}}$ in the stellar disc, ${1.54\times10^{6}}$ in the gas disc, and ${1\times10^{5}}$ in the bulge.
	The dark matter halo particles have a mass of ${9\times10^{5}~\mathrm{M_{\odot}}}$ and the star and gas particle types both have a mass of ${2.7\times10^{3}~\mathrm{M_{\odot}}}$.
	This gives us a ${9\times10^{11}~\mathrm{M_{\odot}}}$ halo, ${1.05\times10^{10}~\mathrm{M_{\odot}}}$ disc (60~per~cent in stars and 40~per~cent in gas), and ${2.7\times10^{8}~\mathrm{M_{\odot}}}$ bulge.

\begin{figure*}
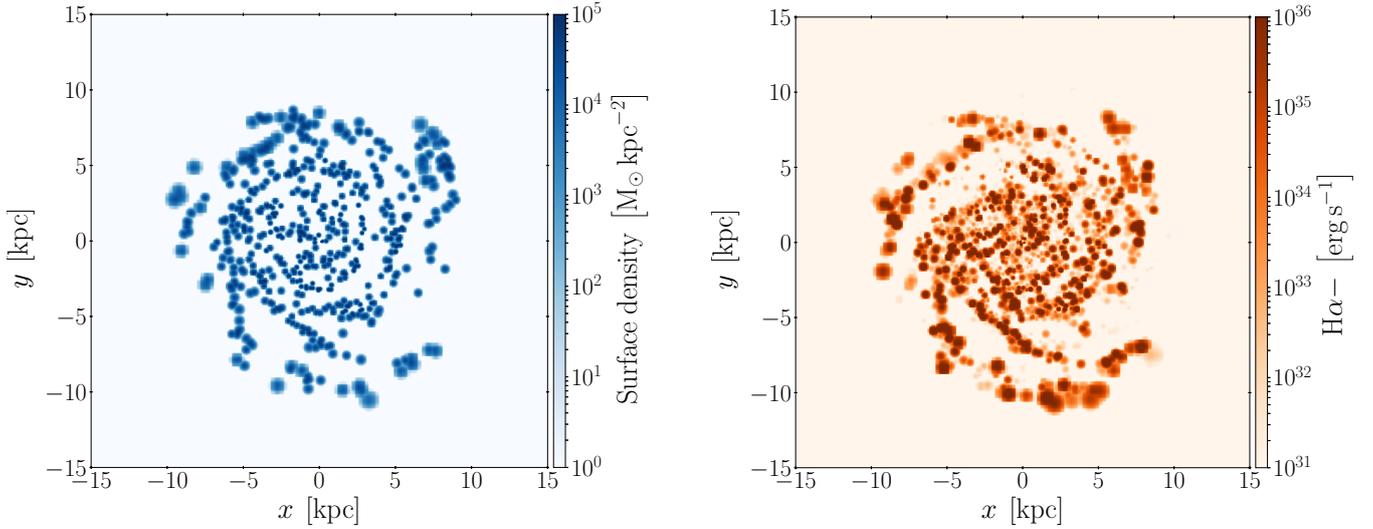

		\centering
		\includegraphics[width=\columnwidth]{\img{QH0_REF}}%
		\hfill%
		\includegraphics[width=\columnwidth]{\img{QH0_TRACER}}%
		\caption{%
			Example maps we use as input for the \textsc{Heisenberg} code.
			\textbf{Left:}~A reference map generated using the mass surface density of star particles in the age range 10--15~Myr, implying a reference time-scale of 5~Myr in this example.
			See \cref{sec:refMap} for details.
			\textbf{Right:}~A synthetic H${\alpha}$ emission map without the continuum (H${\alpha{-}}$) generated by performing SPS on the simulated galaxy.
			See \cref{sec:emissionMap} for details.
		}\label{fig:inputMaps}
	\end{figure*}

	We then evolve the initial conditions for 2.2 Gyr using the smoothed particle hydrodynamics (SPH) code \textsc{P-Gadget-3} \citep[last described by][]{SPRI05b}, which makes use of the \textsc{SPHGal} hydrodynamics solver.
	\textsc{SPHGal} was implemented by \citet{HU14} in order to overcome many of the numerical issues associated with traditional SPH\@.
	To be considered for star formation, gas particles require temperatures less than ${1.2\times 10^{4}}$~K and hydrogen particle densities more than 0.5~${\mathrm{cm^{-3}}}$.
	Stars are formed from eligible gas particles stochastically according to the method described in \citet{KATZ92}.
	Supernova explosions return mass, momentum, and thermal energy back to the ISM\@; these are distributed using a kernel weighting to the 10 nearest gas particles.
	The result of the simulation is a near-${L^{\star}}$ isolated flocculent spiral galaxy, forming stars at a rate of roughly 0.3~${\mathrm{M_{\odot}~yr}^{-1}}$ with a stellar mass of $6.6\times10^9~\mathrm{M_{\odot}}$ and a total cold gas mass of $4.2\times10^9~\mathrm{M_{\odot}}$. These macroscopic galaxy properties are consistent with those of the observed nearby galaxy population, as it resides on the star formation main sequence \citep{SAIN17,CATI18}, with a normal total gas depletion time \citep{BIGI08,LERO13}.
    In addition, \cref{fig:inputMaps} shows a stellar reference map (\cref{sec:refMap}) and a synthetic H${\alpha{-}}$ map (\cref{sec:emissionMap}) of this galaxy, demonstrating that its morphology is similar to that of nearby flocculent spirals like M33 and NGC300.

	The star formation and feedback prescriptions used in the simulation are certainly inadequate to describe the cloud-scale physics governing the evolutionary cycling between molecular gas, star formation, and feedback within galaxies \citep[see e.g.][]{HOPK18,KRUI19,CHEV20c}.
	However, this is not a concern in the context of the problem at hand.
	The goal of this work is not to accurately model cloud-scale star formation and feedback.
	Instead, we aim to determine how quickly SFR tracer emission fades after the formation of a young stellar population, and to do so self-consistently in the context of the \KLPrin{}.
	This can be achieved with any simulation in which (1) the birth sites of star particles approximately conform to a galaxy-like morphology and (2) the formation of young stellar populations proceeds approximately instantaneously.
	
	As shown by the images in \cref{fig:inputMaps}, the former of the above conditions is indeed achieved by the simulation used here. The additional condition that the stellar population within a cloud forms approximately instantaneously is important, because we aim to measure the SFR tracer emission time-scale for a simple stellar population with a single age. In observational applications of the method, it is certainly possible that the young stellar population generating the SFR tracer emission has a non-zero age spread. For this reason, observational applications include an ``overlap'' time-scale that is added to the SFR tracer emission time-scale. This overlap time-scale represents the time for which gas and SFR tracer emission coexist, and is assumed to roughly correspond to the age spread of the stellar population. As a result, the total emission time-scale of the SFR tracer in observations is the sum of the time-scale measured in this paper \emph{and} the overlap time-scale that is observationally inferred with \textsc{Heisenberg}. As discussed in \citet[Sect.~4.3.3]{KRUI18}, the first $1{-}2$ star particles to form in a simulated cloud typically destroy it, such that the cloud-scale star formation in the simulation is effectively instantaneous, as desired.
	
	In summary, the simulation satisfies the above requirements for reliably constraining SFR tracer time-scales. In turn, this will enable observational applications of our method that themselves will motivate a future generation of star formation and feedback models, capable of describing cloud-scale evolutionary cycling in galaxy simulations (see \citealt{KRUI18} and \citealt{FUJI19} for a discussion).

	\subsection{Generation of the reference maps}\label{sec:refMap}
	The role the reference map plays in the \textsc{Heisenberg} code is to calibrate the absolute evolutionary timeline of the star formation process.
	In the context of this paper, it is used to calibrate the characteristic time-scale of the synthetic SFR tracer emission maps.

	In our experiments aimed at measuring the SFR tracer emission time-scales, we need to know the reference time-scale exactly.
	For this reason, we use simulated rather than real galaxies.
	We produce reference maps from the simulation by generating mass surface density maps of the star particles in a specific age bin. This age bin is chosen to cover a relatively narrow and young age interval that is similar in duration to (but temporally offset from) the stellar age interval at which the SFR tracer emission is generated (see below).
	The width of this age bin acts as the reference time-scale, ${t_{\mathrm{R}}}$.
	We smoothen the selected star particles using a Wendland ${C^{4}}$ kernel \citep{DEHN12} (the same kernel \textsc{SPHGal} introduces into \textsc{P-Gadget-3}) over the 200 nearest neighbouring particles; this produces a realistic reference map (i.e.\ not a map of point particles).
	
	As long as the galaxy-average physical conditions do not evolve significantly, and the galaxy is large enough to contain a statistical sample of star-forming regions ($\gtrsim35$, see \citealt{KRUI18}), the above approach provides a reliable reference time-scale. In that case, the number of star-forming regions in a given age interval is simply proportional to the width of that age interval, irrespectively of the absolute age. This is exactly the setup that we require. The reference maps considered in this work contain well over 35 regions (see \cref{fig:inputMaps} and \citealt{KRUI18}) and do not experience any significant macroscopic evolution over the time-scales considered.

	In principle, we have a free choice over the age bin we use.
	However, for the best results and the most realistic set-up there are a few restrictions.
	In \cref{sec:KL14}, we note that \textsc{Heisenberg} is designed such that the reference map corresponds to the second phase of the evolutionary timeline.
	To avoid any overlap between the evolutionary phases, the minimum age of the star particles used in the reference map (${t_{\mathrm{M}}}$) must therefore be at least the duration of the first (SFR tracer emission) phase (${t_{\mathrm{E,\,0}}}$, we include the subscript \enquote{0} to indicate that this is for a well sampled IMF\@: this distinction is necessary in \cref{sec:IMFsampling}) of the evolutionary timeline.
	This defines the lower limit of the stellar age bin used to generate the reference map:
	\begin{equation}\label{eq:Acrit}
		t_{\mathrm{M}} \gtrsim t_{\mathrm{E,\,0}}~\mathrm{.}
	\end{equation}
	This maximises the diagnostic power of the fit by minimising the temporal overlap between both phases, thereby avoiding strongly flattened tuning fork diagrams (see \cref{fig:exampleTuningfork}). At the same time, it is undesirable to select a value of ${t_{\mathrm{M}}}$ much larger than the galactic dynamical time because groups of star particles formed in the same clouds may have dispersed.
	We therefore prefer using ${t_{\mathrm{M}}\approx t_{\mathrm{E,\,0}}}$.

	\begin{figure}
		\centering
		\includegraphics[width=\columnwidth]{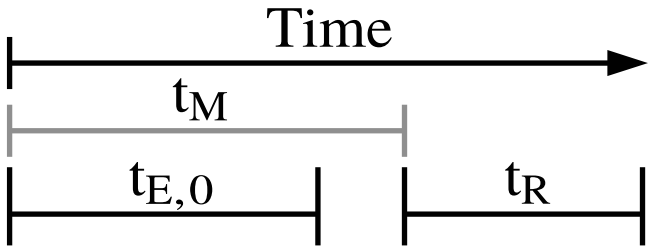}%
		\caption{%
			Schematic diagram showing how the different time-scales we define within the paper are related.
			Time starts at the birth of the star particle.
			The emission map shows the particles formed within a time-scale ${t_{\mathrm{E,\,0}}}$ prior to the simulation snapshot, where ${t_{\mathrm{E,\,0}}}$ represents the characteristic time-scale of the SFR tracer.
			The time over which the reference map runs is defined by ${t_{\mathrm{M}}}$ and ${t_{\mathrm{R}}}$, where ${t_{\mathrm{M}}}$ sets the minimum age of the star particles used to create the reference map and ${t_{\mathrm{R}}}$ defines the width of the age bin and therefore reference time-scale.
			The structure of the \textsc{Heisenberg} code is such that, the duration of second evolutionary phase is used to calibrate the duration of the first.
			This means that to calibrate the SFR tracer time-scale (${t_{\mathrm{E,\,0}}}$), the SFR tracer map must be the first evolutionary phase (i.e.\ ${t_{\mathrm{E,\,0}} \le t_{\mathrm{M}}}$); this is unlike observational applications, where it is usually the second.
		}\label{fig:timeLine}
	\end{figure}

	\citet{KRUI18} show that \textsc{Heisenberg} provides the most accurate measurement of the underlying time-scales if the duration associated to both of the input maps is similar (within a factor of 10, but ideally within a factor of 4).
	This finding sets the preferred width of the age bin:
	\begin{equation}\label{eq:Wcrit}
		t_{\mathrm{R}} \approx t_{\mathrm{E,\,0}}~\mathrm{.}
	\end{equation}
	This maximises the diagnostic power of the fit by favouring similar durations of both phases, thereby avoiding strongly asymmetric tuning fork diagrams (see \cref{fig:exampleTuningfork}). To summarise the above definitions, \cref{fig:timeLine} shows a schematic timeline of how ${t_{\mathrm{M}}}$, ${t_{\mathrm{R}}}$, and ${t_{\mathrm{E,\,0}}}$ are related.

	\begin{table}
		\centering
		\caption{%
			We create input reference maps from the star particles that fall within a particular age bin.
			The age bin, for a given reference map, is defined through ${t_{\mathrm{M}} \leq \mathrm{Age} \leq t_{\mathrm{M}} + t_{\mathrm{R}}}$.
			We show here all the values used in this paper for ${t_{\mathrm{M}}}$ and ${t_{\mathrm{R}}}$ when defining these age bins.
			This results in a ${9 \times 9}$ array of reference maps (see \cref{fig:ageBinTimeScales}).
			See \cref{sec:refMap} for more details.
		}\label{tab:refBin}
		\begin{tabular}{l rrrrrrrrr}
			\toprule
			Emission Type & \multicolumn{9}{c}{${t_{\mathrm{M}}}$ and ${t_{\mathrm{R}}}$ ${\left[\mathrm{Myr}\right]}$}\tabularnewline
			\midrule
			H${\alpha}$ & 1 & 3 & 5 & 7 & 10 & 15 & 20 & 25 & 30\tabularnewline
			UV & 5 & 10 & 15 & 20 & 25 & 30 & 50 & 70 & 100\tabularnewline
			\bottomrule
		\end{tabular}
	\end{table}

	To quantify (and avoid) any systematic biases of the measured SFR tracer time-scale, we investigate the dependence on the choice of stellar age bin used to generate the reference map.
	In practice, this means we vary the values of ${t_{\mathrm{M}}}$ and ${t_{\mathrm{R}}}$.
	We present the range of values we use for ${t_{\mathrm{M}}}$ and ${t_{\mathrm{R}}}$ in \cref{tab:refBin}.
	These are guided by the range of possible characteristic time-scales for H${\alpha}$ and far-UV (FUV) emission found in \citet{LERO12}.
	\citeauthor{LERO12} use the results of \textsc{Starburst99} calculations to determine a characteristic time-scale using several methods: a luminosity-weighted average time, as well as the times at which the tracer emission reaches a particular limit in terms of the total cumulative emission or its instantaneous intensity.

	\subsection{Generation of the emission maps}\label{sec:emissionMap}
	In order to preform our analysis, we need to produce synthetic emission maps of each SFR tracer.
	The simulation that we base this work on (see \cref{sec:simulation}) contains no information about the expected emission spectrum.
	We therefore use \textsc{slug2} \citep{SILV12, SILV14, KRUM15}, a stochastic SPS code, to take the age and mass of the star particles and predict the associated emission for the filters specified in \cref{tab:filterList}.

	With the \textsc{slug2} model, we predict the expected rest-frame emission spectrum for every star particle within the simulation.\footnote{%
		We note that the age binning described in \cref{sec:refMap} is not used for the emission maps.
	}
	The code first samples an IMF to construct a simple stellar population of total mass matching that of the star particle and then uses stellar evolution tracks along with the age of the star particle to determine the combined emission of this simple stellar population.
	\textsc{slug2} then converts the full combined emission spectrum into a single luminosity value for each of the SFR tracers in \cref{tab:filterList} using filter response curves.
	These single luminosity values are what we assign to our star particles when we produce our synthetic rest-frame emission maps.
	We use the same smoothing procedure as we described in \cref{sec:refMap}.
	This means that, even though our star particles are treated as simple stellar populations, the star-forming regions themselves, which are a collection of multiple particles, will have an age spread.
	An example of a synthetic H${\alpha{-}}$ map is shown in \cref{fig:inputMaps}.

	The adopted UV response filters are all included by default in \textsc{slug2} (see \citealt{KRUM15} for more details).
	The H${\alpha}$ SFR tracers, however, require different steps.
	For H${\alpha{-}}$ we use the hydrogen-ionizing photon emission, ${Q\!\left(\mathrm{H^{0}}\right)}$, directly\footnote{%
		A true H${\alpha}$ luminosity can be calculated from ${Q\!\left(\mathrm{H^{0}}\right)}$ using \citet[Equation~2]{SILV14}; however, using the required scaling factor will not change the results we recover here \citep[see][]{KRUI18} and so the conversion is unnecessary.
	} and for ${\mathrm{H\alpha{+}}\,W}$ we define the narrow band filter, ${\mathcal{F}_{\mathrm{H\alpha{+}}\,W}}$, as
	\begin{align}\label{eq:HAFilter}
		\mathcal{F}_{\mathrm{H\alpha{+}}\,W} = \left\lbrace
		\begin{aligned}
			&1 &&6562 -\frac{W}{2}~\text{\AA{}} \leq \lambda \leq 6562 +\frac{W}{2}~\text{\AA{}}
			\\
			&0 &&\mathrm{Otherwise}
		\end{aligned}
		\right.~\mathrm{.}
	\end{align}
	The emission spectrum produced by \textsc{slug2}, includes the H${\alpha}$ emission line but does not calculate the underlying absorption feature from the stellar continuum.
	In \cref{app:HAAbsEmi} we use \textsc{Starburst99} simulations to investigate when the absorption can no longer be neglected.
	We find that for the time-scales we are considering the absorption is negligible.

	For the analysis in \cref{sec:resultsSingleTimeScale}, we use a \citet{CHAB05} IMF with Geneva solar-metallicity 
	evolutionary tracks \citep{SCHA92} and \textsc{Starburst99} spectral synthesis.
	The \textsc{slug2} model samples the IMF non-stochastically\footnote{%
		In \cref{sec:IMFsampling}, we will use the stochastic IMF sampling mode of \textsc{slug2} to investigate its effect on the inferred SFR tracer time-scales.
	}
	(i.e.\ we use a well sampled IMF) and no foreground extinction is applied.
	The surrounding material has a hydrogen number density of ${10^{2}~\mathrm{cm}^{-3}}$.
	We assume that only 73~per~cent of the ionising photons are reprocessed into nebular emission, which is consistent with the estimate from \citet{MCKE97}; this could be because those photons are absorbed by circumstellar dust, or because they escape outside the observational aperture (the observational effects of these two possibilities are indistinguishable).

	We choose to produce our synthetic emission maps without extinction for a number of reasons.
	In observational applications of the \KLPrin{}, there is often some overlap between the first and second phases of the evolutionary timeline.
	For instance, when applying the method to a molecular gas map (e.g.\ CO) and an ionised emission map (e.g.\ H${\alpha}$), there will be some non-zero time for which both tracers coexist.
	When a region resides in this \enquote{overlap} phase, the star-forming region may be partially embedded in dust and gas; during this phase the region suffers the most from extinction.
	We can therefore define the duration of this second phase, ${t_{2}}$, as
	\begin{equation}
		t_{2} = t_{\mathrm{o}} + t_{\mathrm{i}}~\mathrm{,}
	\end{equation}
	where ${t_{\mathrm{o}}}$ is the duration of the second phase that overlaps with the first, and ${t_{\mathrm{i}}}$ the duration that is independent.
	The characteristic time-scales we define in this paper are for this independent part, ${t_{\mathrm{i}}}$, of the second phase.
	This is where the region is no longer embedded in dust and gas and therefore not suffering from significant extinction.
	We motivate this by the notion that molecular gas correlates with star formation: as long as CO emission is present, star formation is likely to be ongoing.
	The \enquote{clock} defined by the SFR tracer lifetime only starts when the last massive stars have formed.
    This does mean that the application of \textsc{Heisenberg} to tracers other than CO may require a different definition of the reference time-scale.
	To facilitate this, the \textsc{Heisenberg} code enables the user to specify if the reference time-scale includes or excludes this overlap phase (see \citealt[Section~3.2.1]{KRUI18}).

	In addition, it is desirable to exclude extinction for two further reasons.
	Firstly, the effects of extinction can, in most cases, be significantly reduced if not completely corrected for (e.g.\ \citealt{JAME05}), meaning in practice the input maps provided to \textsc{Heisenberg} can be corrected for extinction.
	Secondly, if we perform our analysis with extincted maps, the results would no longer be generally applicable and would only apply to galaxies that suffer from the same amount of extinction.
	Our current approach therefore enables constructing a \enquote{universal} baseline of extinction-corrected SFR tracer lifetimes.
	In future work, we aim to consider extinction using galaxy simulations covering a range of gas surface densities \citep{HAYD19}.

	\section{Characteristic time-scales for a fully sampled IMF at solar metallicity}\label{sec:resultsSingleTimeScale}
\begin{figure*}
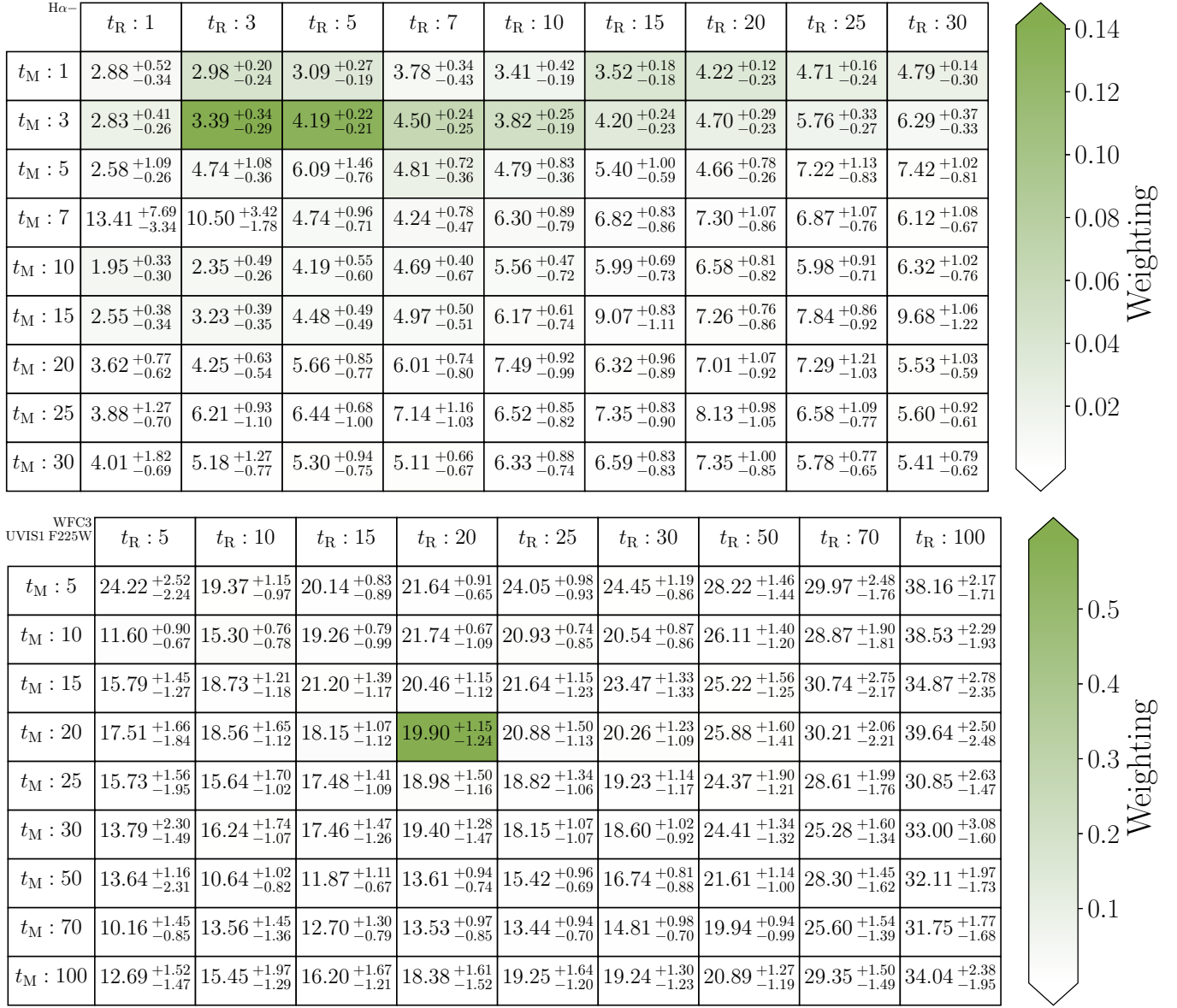

		\centering
		\includegraphics[width=\textwidth]{\img{modc020_QH0_weights}}%
		\\[\baselineskip]
		\includegraphics[width=\textwidth]{\img{modc020_WFC3_UVIS_F225W_weights}}%
		\caption{%
			Two examples showing the range of characteristic time-scale values (and associated uncertainties) as determined using \textsc{Heisenberg} for different reference maps.
			\textbf{Top:}~H${\alpha}$ emission excluding the continuum (H${\alpha{-}}$).
			\textbf{Bottom:}~UV emission (WFC3 UVIS F225W).
			The reference maps are characterised by the age bin used to select the star particles which are included in the reference map.
			${t_{\mathrm{M}}}$ denotes the minimum age of the star particles and ${t_{\mathrm{R}}}$ the width of the age bin.
			The colour-coding is based on the weighting, ${\mathcal{W}}$, used when calculating the weighted average.
			All values within the tables are given in Myr.
		}\label{fig:ageBinTimeScales}
	\end{figure*}

	We constrain the characteristic time-scales for several SFR tracers by applying the \textsc{Heisenberg} code to the synthetic SFR tracer maps and reference maps described in \cref{sec:methodNonStoc}.
	The reference maps show the star particles in a chosen age interval. Since there is some freedom in choosing this interval,  we measure the SFR tracer time-scale for a wide variety of age intervals rather than picking a single one. \cref{app:weights} demonstrates this improves the accuracy of the measurements relative to using a single age interval. Mathematically, we change the values of ${t_{\mathrm{M}}}$ and ${t_{\mathrm{R}}}$, which define the age interval as ${t_{\mathrm{M}} \leq \mathrm{Age} \leq t_{\mathrm{M}}+t_{\mathrm{R}}}$.
	This approach generates an array of SFR tracer time-scales, spanned by ${t_{\mathrm{M}}}$ and ${t_{\mathrm{R}}}$.
	We now first describe how we reduce these \enquote{time-scale arrays} (see \cref{fig:ageBinTimeScales} for examples) into a single characteristic time-scale for each SFR tracer. To achieve this, we use the complete probability density function (PDF) of each measured time-scale, which is provided by \textsc{Heisenberg}.

    \subsection{Mathematical procedure} \label{sec:weights}
	In principle, \textsc{Heisenberg} allows any pair of values to be used for generating the reference map, but the accuracy of the method is highest when the durations of both evolutionary phases (${t_{\mathrm{E,\,0}}}$ and ${t_{\mathrm{R}}}$) are similar \citep{KRUI18}.
	In that case, the tuning fork diagram of \cref{fig:exampleTuningfork} is symmetric, allowing the method to retrieve the underlying time-scales with a precision of better than 30~per~cent.
	We additionally prefer numerical experiments with a minimal time offset between the reference map and the SFR tracer, to mimic the close evolutionary correspondence between molecular gas and (massive) star formation. We therefore also prefer solutions in which the temporal offset of the reference phase (${t_{\mathrm{M}}}$) is similar to the duration of the SFR tracer (${t_{\mathrm{E,\,0}}}$).
	Since we do not know the latter a priori, it is not possible to choose an optimal pair of ${t_{\mathrm{R}}}$ and ${t_{\mathrm{M}}}$ in advance.
	
	In order to condense an array of SFR tracer time-scale measurements (e.g.\ \autoref{fig:ageBinTimeScales}) into a single characteristic time-scale, we must design a simple weighting scheme that accounts for the above behaviour of \textsc{Heisenberg}. This scheme should favour solutions with  ${t_{\mathrm{M}}\sim t_{\mathrm{R}}\sim t_{\mathrm{E,\,0}}}$. Additionally, it should weigh each measurement by the inverse-square of its uncertainty, as is common when calculating the average across a sample of measurements.

	First, we weight each measurement simply by the inverse of its geometric distance from ${t_{\mathrm{R}}}$ and ${t_{\mathrm{M}}}$ in logarithmic space:
	\begin{equation}\label{eq:distance}
		\mathcal{W}^{\mathrm{d}}_{ij} = {\left\{{\left[\log_{10}\left(\frac{t_{ij}}{t_{\mathrm{M},\,i}}\right)\right]}^{2} + {\left[\log_{10}\left(\frac{t_{ij}}{t_{\mathrm{R},\,j}}\right)\right]}^{2}\right\}}^{-\frac{1}{2}}~\mathrm{.}
	\end{equation}
	This is the simplest possible approach to performing a proximity-based weighting. It favours more strongly elements that satisfy the criteria we describe in \cref{eq:Acrit,eq:Wcrit} (i.e.\ the closer ${t_{ij}}$ is to ${t_{\mathrm{M},\,i}}$ and ${t_{\mathrm{R},\,j}}$, the better).
	
	Secondly, we weight each measurement by the inverse-square of its uncertainty, which accounts for asymmetric uncertainties by taking the mean of the lower and upper uncertainty:\footnote{%
		Using the average of the lower and upper uncertainty is not technically correct; however, the methods as suggested by \citet{BARL03} would have little impact on the final result and so are neglected.
	}
	\begin{equation}\label{eq:uncertainty}
		\mathcal{W}^{\mathrm{u}}_{ij} = {\left(\frac{\sigma^-_{ij} + \sigma^{+}_{ij}}{2}\right)}^{-2}~\mathrm{.}
	\end{equation}
	The weights ${\mathcal{W}^{\mathrm{d}}}$ and ${\mathcal{W}^{\mathrm{u}}}$ are then combined and normalised as:
	\begin{equation}\label{eq:weight}
		\mathcal{W}_{ij} = \frac{\mathcal{W}^{\mathrm{d}}_{ij} \mathcal{W}^{\mathrm{u}}_{ij}}{\sum_{ij}\mathcal{W}^{\mathrm{d}}_{ij} \mathcal{W}^{\mathrm{u}}_{ij}}~\mathrm{.}
	\end{equation}

	The above weighting scheme appropriately combines all measurements in the time-scale array into a single SFR tracer time-scale. However, it does not propagate the individual measurement uncertainties into the final measurement. To accomplish this, we adopt a simple Monte-Carlo approach. We produce ${10^{6}}$ realisations of the time-scale array, where the value of each element of each realisation of the time-scale array has been randomly sampled from its associated PDF. For each of the ${10^{6}}$ realisations of the time-scale array we calculate the weighted mean according to \cref{eq:weight,eq:distance,eq:uncertainty}. This process results in ${10^{6}}$ characteristic time-scales, from which we take the median to define the characteristic time-scale and the 16th and 84th percentiles to define the uncertainties. These uncertainties contain both the uncertainties on the individual measurements and the systematic uncertainty associated with choosing a combination of ${t_{\mathrm{M}}}$ and ${t_{\mathrm{R}}}$.

	With the above procedure, we condense the array of time-scales into a single number that most strongly weighs the values that are the most accurate \citep[based on][]{KRUI18} and those that have the smallest measurement uncertainties.
	In practice, we find that the typical standard deviation of all SFR tracer time-scales is $\sim0.15$~dex, over a dynamical range of 1.5~dex in ${t_{\mathrm{M}}}$ and ${t_{\mathrm{R}}}$. This demonstrates that the inferred time-scales are not extremely sensitive to the choice of reference map, but the full array of reference maps does allow us to optimise the accuracy of the SFR tracer time-scales.
	In \cref{app:weights}, we demonstrate that our weighting scheme indeed performs better than the approach of choosing a single measurement by simply minimising the difference in geometric distance between ${t_{\mathrm{E,\,0}}}$, ${t_{\mathrm{M}}}$, and ${t_{\mathrm{R}}}$ (or maximising $\mathcal{W}^{\mathrm{d}}_{ij}$).

	\subsection{Measured SFR tracer time-scales}
	\begin{figure*}
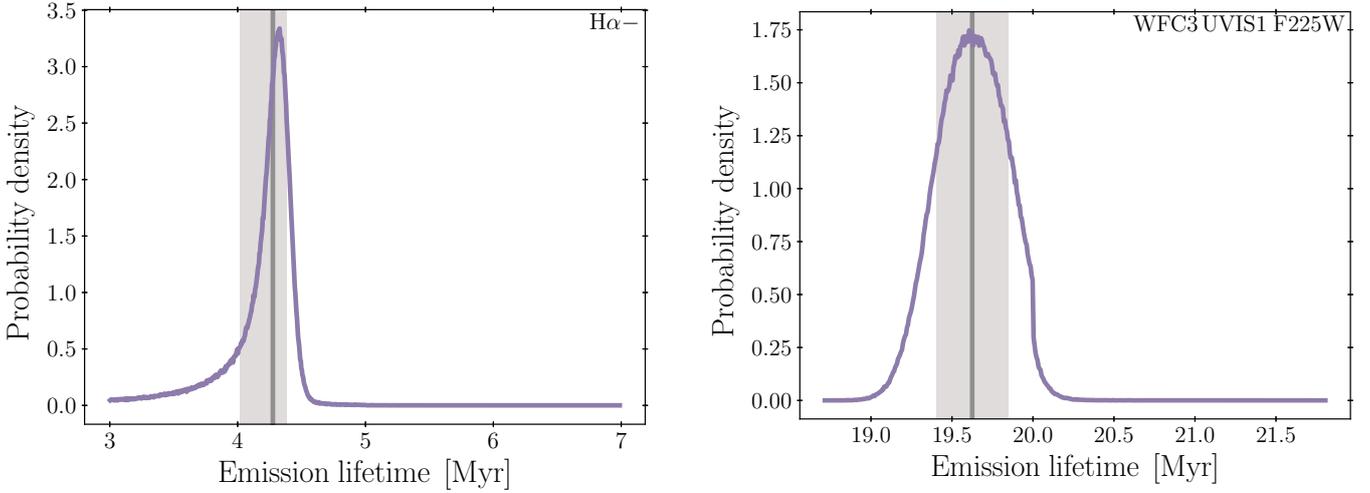

		\centering
		\VerticalCentered{\includegraphics[width=\columnwidth]{\img{modc020-QH0_dist}}}%
		\hfill%
		\VerticalCentered{\includegraphics[width=\columnwidth]{\img{modc020-WFC3_UVIS_F225W_dist}}}%
		\caption{%
			Two examples showing the probability density functions associated to the defined characteristic time-scales.
			\textbf{Left:}~H${\alpha}$ emission excluding the continuum (H${\alpha{-}}$).
			\textbf{Right:}~UV emission (WFC3 UVIS F225W).
			The vertical line shows the selected time-scale (the median of the distribution); the shaded region, the uncertainty defined by the 16th and 84th percentiles.
		}\label{fig:TSPDF}
	\end{figure*}
	When applying \textsc{Heisenberg} to the pairs of reference and SFR tracer maps, we use the default input parameters specified in Tables~1 and~2 of \citet{KRUI18}.
	The only exceptions are as follows.
	We set \texttt{tstar\_incl} = 1, to indicate that the reference time-scale (i.e.\ the width of the age bin) also includes the overlapping phase.\footnote{%
		At first glance, this may seem to contradict the discussion in \cref{sec:emissionMap}, but this is not the case. In \cref{sec:emissionMap}, we explain that the characteristic time-scales of the SFR tracers we define do not include the overlap phase; and so, when using the characteristic time-scales we present here, one should use \texttt{tstar\_incl} = 0. The analysis we perform to define the characteristic time-scales, uses a reference map produced from star particles in a specific age bin. The width of this age bin is used as the reference time-scale and is the total duration of that phase: this includes any overlap.
	}
	As we are not making any cuts in galactocentric radius, we also set \texttt{cut\_radius} = 0.
	Finally, we define the range of aperture sizes using a minimum aperture size of ${l_{\mathrm{ap,}\,\min} = 25}$~pc and a number of ${N_{\mathrm{ap}} = 17}$ apertures, to produce 17 logarithmically-spaced aperture diameters from 25--6400~pc.

	In \cref{fig:ageBinTimeScales}, we present the time-scale arrays obtained for H${\alpha{-}}$ and WFC3 UVIS F225W SFR tracers.
	These time-scale arrays only serve as examples, since the elements show the output of \textsc{Heisenberg} and are not from the ${10^{6}}$ Monte-Carlo realisations.
	\cref{fig:TSPDF} shows the PDFs of the defined characteristic time-scale for H${\alpha{-}}$ and WFC3 UVIS F225W, as obtained by applying the weighting scheme described in \cref{sec:weights} to $10^6$ different Monte-Carlo realisations of \cref{fig:ageBinTimeScales}.
	
	Reassuringly, the best-fitting numbers in \cref{fig:ageBinTimeScales} show relatively limited cell-to-cell variation, which means that stochastic SFR variations in the simulation (which would affect different age bins differently) do not strongly affect our results. This is to be expected -- \citet{KRUI18} show that the SFR varies by less than 30~per~cent over the typical age intervals considered. This may not always hold in observational applications of our methodology, and should therefore not be applied to systems with strongly varying SFRs, such as galaxy mergers or other starbursts (also see Sects.~4.2.4.3 and 4.4 of \citealt{KRUI18}).
	
	\cref{tab:resultsSolar} lists the characteristic time-scales and associated uncertainties obtained for each of the different SFR tracers. The complete set of SFR tracer time-scales spans a range of $4{-}34$~Myr. We use these measurements to define age bins that we will use to generate reference maps when investigating the impact of metallicity and IMF sampling in later sections. These are listed in the final column of the table and are calculated as $t_{\mathrm{E,\,0}} \le \mathrm{Age} \le 2t_{\mathrm{E,\,0}}$.

	\begin{table}
		\centering
		\caption{%
			The characteristic time-scales, ${t_{\mathrm{E,\,0}}}$, obtained for the different SFR tracers (see \cref{tab:filterList} for details) and the corresponding age bins (${t_{\mathrm{E,\,0}} \le \mathrm{Age} \le 2t_{\mathrm{E,\,0}}}$) for producing reference maps in later sections of this paper.
			These results are for a well sampled IMF at solar metallicity.
			The filter order is in increasing filter width (${W}$) for H${\alpha{+}}$ and increasing response-weighted mean wavelength  (${\overline{\lambda}_{\mathrm{w}}}$) for UV\@.
			This table is an extract of \cref{tab:resultsAllZ}, which includes the characteristic time-scales and age bins for different stellar metallicities (${Z/\mathrm{Z_{\odot}} = \text{0.05--2}}$).
		}\label{tab:resultsSolar}
		\begin{tabular}{lrr}
			\toprule
			{} & ${t_{\mathrm{E,\,0}}}$ ${\left[\mathrm{Myr}\right]}$ & Age bin ${\left[\mathrm{Myr}\right]}$\tabularnewline
			\midrule
			H${\alpha{-}}$ & ${4.3^{+0.1}_{-0.3}}$ & 4.3--8.6\tabularnewline
			H${\alpha{+}}$ 10 \AA{} & ${5.6^{+0.2}_{-0.1}}$ & 5.6--11.1\tabularnewline
			H${\alpha{+}}$ 20 \AA{} & ${7.3^{+0.4}_{-0.2}}$ & 7.3--14.6\tabularnewline
			H${\alpha{+}}$ 40 \AA{} & ${9.3^{+0.2}_{-0.3}}$ & 9.3--18.6\tabularnewline
			H${\alpha{+}}$ 80 \AA{} & ${10.7^{+0.2}_{-0.2}}$ & 10.7--21.4\tabularnewline
			H${\alpha{+}}$ 160 \AA{} & ${16.4^{+0.6}_{-0.3}}$ & 16.4--32.7\tabularnewline
			GALEX FUV & ${17.1^{+0.4}_{-0.2}}$ & 17.1--34.2\tabularnewline
			UVOT W2 & ${19.0^{+0.3}_{-0.2}}$ & 19.0--38.0\tabularnewline
			WFC3 UVIS1 F218W & ${19.4^{+0.2}_{-0.2}}$ & 19.4--38.9\tabularnewline
			UVOT M2 & ${19.5^{+0.2}_{-0.2}}$ & 19.5--39.0\tabularnewline
			GALEX NUV & ${19.6^{+0.2}_{-0.2}}$ & 19.6--39.1\tabularnewline
			WFC3 UVIS1 F225W & ${19.6^{+0.2}_{-0.2}}$ & 19.6--39.3\tabularnewline
			WFPC2 F255W & ${22.4^{+0.2}_{-0.2}}$ & 22.4--44.7\tabularnewline
			UVOT W1 & ${21.8^{+0.2}_{-0.2}}$ & 21.8--43.5\tabularnewline
			WFC3 UVIS1 F275W & ${23.5^{+0.2}_{-0.2}}$ & 23.5--47.0\tabularnewline
			WFPC2 F300W & ${27.7^{+0.6}_{-0.3}}$ & 27.7--55.4\tabularnewline
			WFPC2 F336W & ${33.1^{+0.4}_{-0.3}}$ & 33.1--66.3\tabularnewline
			WFC3 UVIS1 F336W & ${33.3^{+0.4}_{-0.4}}$ & 33.3--66.6\tabularnewline
			\bottomrule
		\end{tabular}
	\end{table}

\begin{figure}
		\centering
		\includegraphics[width=\columnwidth]{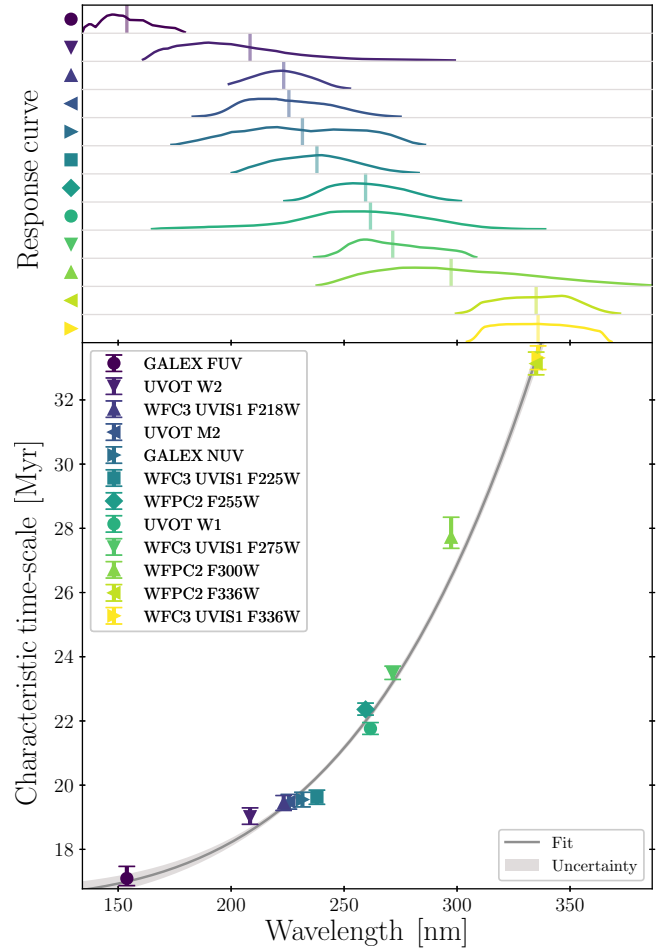}%
		\caption{%
			\textbf{Top:}~The normalised response curves of the UV filters considered in this paper (also see \cref{tab:UVfilterList}).
            The vertical lines indicate the response-weighted mean wavelengths, ${\overline{\lambda}_{\mathrm{w}}}$.
			\textbf{Bottom:}~Characteristic emission time-scales for UV filters as a function of response-weighted mean wavelength.
            The grey curve shows the fit described in \cref{eq:UVtimeSolarFit} and the shaded regions indicates the associated uncertainty.
		}\label{fig:uvTimeScales}
	\end{figure}

	\cref{fig:uvTimeScales} shows the UV-based SFR tracer time-scales as a function of the response curve-weighted mean wavelength, ${\overline{\lambda}_{\mathrm{w}}}$. The figure shows both are closely correlated, such that similar response-weighted mean wavelengths give similar characteristic time-scales.
	We perform a weighted least-squares minimization to obtain a relation between ${\overline{\lambda}_{\mathrm{w}}}$ and the UV characteristic time-scale, ${t_{\mathrm{E,\,0}}^{\mathrm{UV}}}$:
	\begin{equation}\label{eq:UVtimeSolarFit}
		t_{\mathrm{E,\,0}}^{\mathrm{UV}}~\left[\mathrm{Myr}\right] =  \left(3.00^{+0.29}_{-0.31}\right) {\left(\frac{\overline{\lambda}_{\mathrm{w}}}{225~\mathrm{nm}}\right)}^{\left(4.34^{+0.24}_{-0.20}\right)} + \left(16.42^{+0.36}_{-0.31}\right)~\mathrm{.}
	\end{equation}
	The uncertainties on the parameters are calculated using a Monte Carlo approach.
	This analytical expression enables finding the characteristic emission time-scales for UV filters other than the specific ones that we have considered here.

\begin{figure}
		\centering
		\includegraphics[width=\columnwidth]{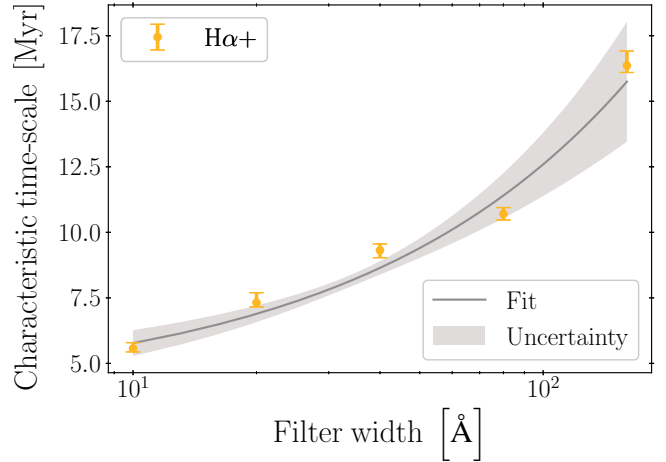}%
		\caption{%
			Characteristic emission time-scales for H${\alpha{+}}$ filters as a function of the filter width.
            The grey curve shows the fit described in \cref{eq:HA+timeSolarFit} and the shaded region indicates the associated uncertainty.
		}\label{fig:HATimeScales}
	\end{figure}

    Similarly, we derive a relation between the H${\alpha{+}}$  characteristic time-scale and filter width:
	\begin{equation}\label{eq:HA+timeSolarFit}
		t_{\mathrm{E,\,0}}^{\mathrm{H\alpha{+}}}~\left[\mathrm{Myr}\right] = \left(4.8^{+1.3}_{-1.3}\right) {\left(\frac{W}{40~\text{\AA{}}}\right)}^{\left(0.65^{+0.20}_{-0.13}\right)} + \left(3.8^{+1.1}_{-1.1}\right)~\mathrm{.}
	\end{equation}
	This relation is compared to the measurements in \cref{fig:HATimeScales}.
	Note that the increase in characteristic time-scale with filter width is not due to a change in the H${\alpha}$ emission, but results from a change in flux from the long-lived continuum emission.

	The characteristic time-scales that we recover (4.3--16.4~Myr for H$\alpha$; 17.1--33.3~Myr for UV) fall within the ranges often quoted in the literature (2--10 Myr and 10--50~Myr, respectively \citealt{KENN12,LERO12}).
	In part, the large variation in literature values reflects the broad range of criteria used for defining the characteristic time-scale of an SFR tracer.
	With the approach taken in this paper, we have remedied this problem for future observational applications of the \KLPrin{} by adopting a specific definition of the SFR tracer time-scale.

	The heterogeneous situation in the previous literature is nicely illustrated by \citet{LERO12}, who present a table of characteristic time-scales for H${\alpha}$ and FUV (at 150~nm).
	Multiple time-scales are listed for each SFR tracer; these time-scales are defined by the duration required to reach a given percentage (50 or 95~per~cent) of the cumulative emission, or of the emission intensity at 1~Myr. These choices are arbitrary, but it is reasonable to ask whether any single percentage of the 1~Myr intensity or cumulative emission can be defined that would correspond to our measured SFR tracer time-scales. To verify this, we take the emission evolution from \citet[Figure 1]{LERO12} and find which percentages correspond to the characteristic emission time-scales we determine for H${\alpha{-}}$ and GALEX FUV. We list these percentage limits in \cref{tab:comparison}. We find that no single percentage limit corresponds to the measured time-scales.\footnote{%
		This also holds when we perform the analysis for the other metallicities considered: ${Z/\mathrm{Z_{\odot}} = 0.05,~0.20,~0.40,~2.00}$ (see \cref{sec:Metallicity} for more details).
		There is also no consistent percentage for a single tracer across the metallicity range.
	}
	As there is no consistent limit, the characteristic time-scale for each SFR tracer must be determined individually. This further validates the approach taken in this study.
	\begin{table}
		\centering
		\caption{%
			Percentages of the emission intensity relative to its instantaneous value at 1~Myr and of its cumulative value over 100~Myr, evaluated at the characteristic time-scales of the SFR tracers presented in \cref{tab:resultsSolar}, based on Figure 1 of \citet{LERO12}.%
		}\label{tab:comparison}
		\TableFootnote{%
			\begin{tabular}{lrr}
				\toprule
				& H${\alpha{-}}$ ${\left[\%\right]}$ & FUV\textsuperscript{a} ${\left[\%\right]}$\tabularnewline
				\midrule
				\% of intensity at 1~Myr & ${19.4^{+2.8}_{-0.7}}$ & ${8.6^{+0.1}_{-0.2}}$\tabularnewline
				\%  of cumulative emission & ${92.4^{+0.6}_{-1.5}}$ & ${76.5^{+0.3}_{-0.2}}$\tabularnewline
				\bottomrule
			\end{tabular}%
		}{\textsuperscript{a} GALEX FUV}
	\end{table}

	In summary, our SFR tracer time-scales fall in the range of commonly reported literature values. These do not correspond to any fixed percentage of the initial or cumulative emission in each tracer. For this reason, each SFR tracer time-scale must be determined individually using the presented method.
	We provide analytic functions (see \cref{eq:UVtimeSolarFit,eq:HA+timeSolarFit}) relating the characteristic emission time-scales for UV and H${\alpha{+}}$ filters to their filter properties, allowing our results to be extended to any UV or H${\alpha{+}}$ filter.

	\section{The effects of metallicity}\label{sec:Metallicity}
	So far, we have only considered stellar populations of solar metallicity. However, it is well-known that the metallicity affects stellar lifetimes \citep[e.g.][]{LEIT99} and thus the characteristic emission time-scales of SFR tracers.
	In order to facilitate observational applications of the \KLPrin{} to the broadest possible range of galaxies, we therefore quantify how the SFR tracer time-scales depend on metallicity.
	In this section, we repeat the experiments performed in \cref{sec:resultsSingleTimeScale} but this time we produce synthetic SFR tracer emission maps using evolutionary tracks of metallicities ${Z/\mathrm{Z}_{\odot} = 0.05,~0.20,~0.40,~2.00}$ \citep{SCHA92, CHAR93, SCHA93, SCHA93b}.

\begin{figure}
		\centering
		\includegraphics[width=\columnwidth]{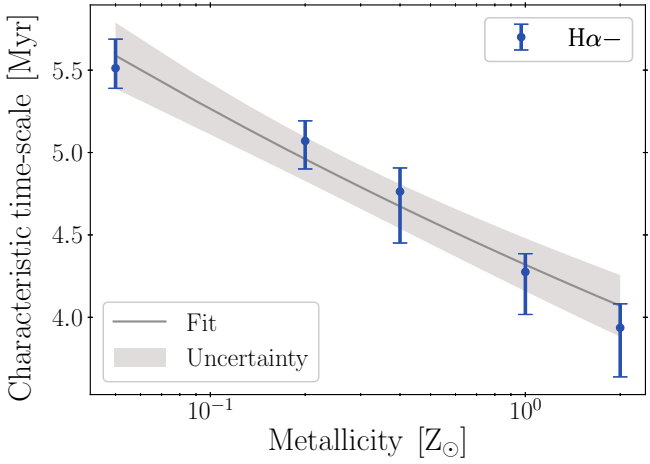}%
		\caption{%
			The relation between metallicity and characteristic time-scale for a well-sampled IMF for H${\alpha{-}}$ filters.
            The grey curve gives the fits described by \cref{eq:HAtimeMetalFit}.
            The shaded region indicates the associated uncertainty.
		}\label{fig:metalTimescale}
	\end{figure}

\begin{figure}
		\centering
		\includegraphics[width=\columnwidth]{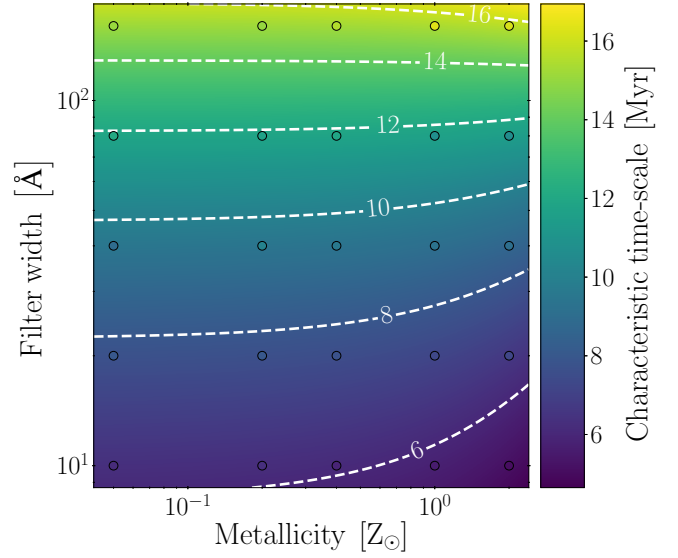}%
		\caption{%
			The surface described by \cref{eq:HAWidtimeMetalFit}, which relates the metallicity and filter width, ${W}$, of a H${\alpha{+}}$ filter to the associated characteristic time-scale for a well sampled IMF\@.
            The data points show the measurements coloured using the same colour bar.
            The surface fits best when it matches the colour of the data points.
		}\label{fig:ZWid_surface}
	\end{figure}

\begin{figure}
		\centering
		\includegraphics[width=\columnwidth]{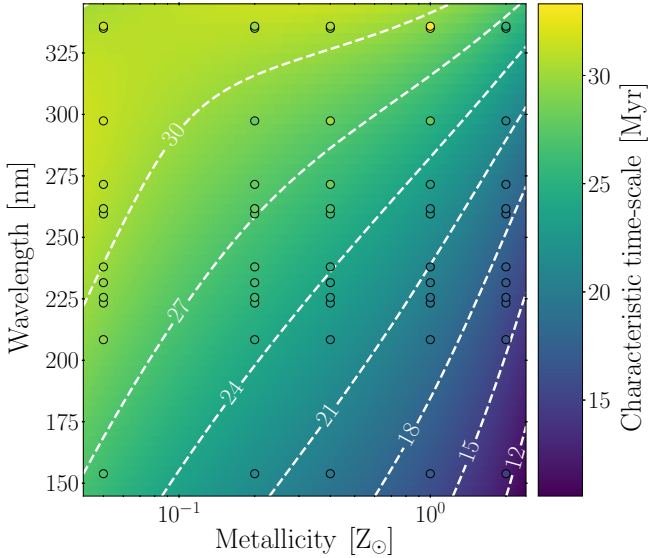}%
		\caption{%
			The surface described by \cref{eq:UVtimeMetalFit}, which relates the metallicity and response-weighted mean wavelength, ${\overline{\lambda}_{\mathrm{w}}}$, of a UV SFR tracer to the associated characteristic time-scale for a well sampled IMF\@.
            The data points show the measurements coloured using the same colour bar.
            The surface fits best when it matches the colour of the data points.
		}\label{fig:ZW_surface}
	\end{figure}

	In \cref{app:completeTable}, we list the resulting characteristic time-scales for a well sampled IMF for all metallicities (also including the solar metallicity results from \cref{tab:resultsSolar}) and the age bins we select for producing reference maps.
	We show the ${Z - t_{\mathrm{E,\,0}}}$ relation in \cref{fig:metalTimescale} for H${\alpha{-}}$, in \cref{fig:ZWid_surface} for H${\alpha{+}}$ filters, and in \cref{fig:ZW_surface} for the UV filters. For all tracers, we find that the characteristic time-scale decreases with metallicity.
	We also include empirical fits described by
	\begin{equation}\label{eq:HAtimeMetalFit}
		t_{\mathrm{E,\,0}}^{\mathrm{H\alpha{-}}}~\left[\mathrm{Myr}\right]= \left(4.32^{+0.09}_{-0.23}\right)  {\left(\frac{Z}{\mathrm{Z_{\odot}}}\right)}^{\left(-0.086^{+0.010}_{-0.023}\right)}~\mathrm{,}
	\end{equation}
	for H${\alpha{-}}$,
	\begin{equation}\label{eq:HAWidtimeMetalFit}
		\begin{split}
			t_{\mathrm{E,\,0}}^{\mathrm{H\alpha{+}}}~\left[\mathrm{Myr}\right] & =
			\left(8.98^{+0.40}_{-0.50}\right) {W_{0}}^{\left(0.265^{+0.028}_{-0.051}\right)}\\
			& + \left(0.23^{+0.15}_{-0.11}\right) Z_{0} W_{0} \\
			& - \left(0.66^{+0.12}_{-0.19}\right) Z_{0} + \left(0.55^{+0.46}_{-0.29}\right) W_{0}~\mathrm{,}
		\end{split}
	\end{equation}
	for H${\alpha{+}}$, and
	\begin{equation}\label{eq:UVtimeMetalFit}
		\begin{split}
			t_{\mathrm{E,\,0}}^{\mathrm{UV}}~\left[\mathrm{Myr}\right] & = -
			\left(0.40^{+0.11}_{-0.16}\right) Z_{1} \lambda_{1} \\
			& + \left(4.5^{+1.3}_{-0.9}\right) Z_{1} + \left(0.70^{+0.26}_{-0.18}\right) \lambda_{1}\\
			& - \left(3.11^{+0.14}_{-0.13}\right) Z_{0} + \left(10.98^{+0.46}_{-0.48}\right) \lambda_{0}\\
			& + \left(7.6^{+1.2}_{-1.6}\right)~\mathrm{,}
		\end{split}
	\end{equation}
	for the UV filters, where
	\begin{equation}
		\begin{split}
			\begin{aligned}
				Z_{0} & \equiv \frac{Z}{\mathrm{Z_{\odot}}}~\mathrm{;} &
				\lambda_{0} &\equiv \frac{\overline{\lambda}_{\mathrm{w}}}{225~\mathrm{nm}}\mathrm{;} &
				W_{0} & \equiv \frac{W}{40~\mathrm{\text{\AA{}}}}\mathrm{;} &
			\end{aligned}\\
			\begin{aligned}
				Z_{1} & \equiv Z_{0}^{\left(-0.313^{+0.051}_{-0.048}\right)}~\mathrm{;} &
				\lambda_{1} & \equiv \lambda_{0}^{\left(6.52^{+0.73}_{-0.71}\right)}~\mathrm{.} &
			\end{aligned}
		\end{split}
	\end{equation}
	As before, we determine the free parameters using a weighted least-squares minimization and the uncertainties through Monte Carlo methods.
	With these relations, it is straightforward to recover the characteristic time-scale for any combination of metallicity and filter properties, without needing to repeat the analysis of this paper.

	\cref{fig:metalTimescale} shows that the characteristic time-scales of H${\alpha{-}}$ change by less than 2~Myr over the metallicity range ${\left[0.05~Z_{\mathrm{\odot}},~2~Z_{\mathrm{\odot}}\right]}$.
	The ranges of characteristic time-scales (3.9--5.5~Myr for H${\alpha{-}}$) fall within the range of literature values (1.7--10~Myr, \citealt{KENN12,LERO12}).

	In \cref{sec:resultsSingleTimeScale}, we describe a curve which relates the filter width, ${W}$, to the characteristic time-scale of ${\mathrm{H \alpha{+}}}$ filters, ${t_{\mathrm{E,\,0}}^{\mathrm{H\alpha{+}}}}$, at solar metallicity.
	\cref{eq:HAWidtimeMetalFit} now extends this relation to include different metallicities, producing a surface in (${t_{\mathrm{E,\,0}}^{\mathrm{H\alpha{+}}},~W,~Z}$) space.
	As mentioned in \cref{sec:resultsSingleTimeScale}, the measured ${\mathrm{H \alpha{+}}}$ time-scales reside at the higher end of (and partially exceeds) the literature range of H$\alpha$ time-scales, because wider filters include more of the long-lived continuum emission.

	Analogously to ${t_{\mathrm{E,\,0}}^{\mathrm{H\alpha{+}}}}$, we can extend the relation given for ${t_{\mathrm{E,\,0}}^{\mathrm{UV}}}$ as a function of response-weighted mean wavelength (\cref{eq:UVtimeSolarFit}) to also include metallicity.
	We obtain a good fit, with the strongest deviations arising at long (${\lambda_{\mathrm{w}}>290}$~nm) wavelengths.
	For UV filters at these wavelengths, we recommend interpolating the data points (provided in \cref{app:completeTable}) rather than adopting \cref{eq:UVtimeMetalFit}.
	The range of characteristic time-scales found for the UV filters (14.5--33.3~Myr) again fall within the range quoted in literature (10--100~Myr, \citealt{KENN12,LERO12}), but tend towards the low end of this range.
	This is a direct result of the fact that the UV emission from star-forming regions fades with time, and the measured time-scales are naturally biased to the ages of regions from which most UV photons emerge.
	
	In \cref{app:fits}, we additionally provide figures showing the one-dimensional projections of the data and fits. These show slices across the two-dimensional distributions shown in \cref{fig:ZWid_surface} and \cref{fig:ZW_surface} for the H${\alpha{+}}$ and UV filters, respectively.
    We include a projection for each metallicity and each equivalent width or response-weighted mean wavelength and use the fits given in \cref{eq:HAWidtimeMetalFit} and \cref{eq:UVtimeMetalFit}. 
	These additional figures enable a direct assessment of how well the fits describe the characteristic time-scales at a specific metallicity or wavelength.

	In summary, we see that the characteristic time-scales decrease with increasing metallicity.
	Observational applications of \textsc{Heisenberg} should therefore use an SFR tracer time-scale appropriate for the metallicity of the observed region.
	We define empirical relations between the SFR tracer time-scale and the metallicity (for H${\alpha{-}}$, \cref{eq:HAtimeMetalFit}) and the filter properties (for ${\mathrm{H \alpha{+}}}$ and UV filters, \cref{eq:HAWidtimeMetalFit,eq:UVtimeMetalFit}).
	For ${\mathrm{H \alpha{+}}}$ and UV SFR tracers, these relations enable the definition of time-scales even for filters that are not explicitly considered here.

	\section{The effects of IMF sampling}\label{sec:IMFsampling}
	In the previous sections, we determine the characteristic time-scales of SFR tracers using synthetic emission maps where \textsc{slug2} fully samples the IMF\@.
	In observational applications of the \KLPrin{}, there is no guarantee (or requirement from \textsc{Heisenberg}) that the regions under consideration have a well sampled IMF\@.
	It is therefore important to investigate the impacts of stochastic IMF sampling on the characteristic time-scales of the SFR tracers, in particular for low-mass star forming regions.

	We describe in \cref{sec:KL14} how the abundance of regions in each input map reflects the duration associated to that map. If the IMF is not well-sampled, the SFR tracer flux is reduced. In the extreme case, a region might not be able to form stars of sufficient mass to produce the SFR tracer emission at all, and would thus be invisible in that tracer.
	These effects are particularly important for the ${\mathrm{H\alpha{\pm}}}$ filters, as H${\alpha}$ emission requires high mass stars (${> 8~\mathrm{M_{\odot}}}$) and is dominated by stars of even higher masses.
	We therefore expect that as the IMF becomes less well-sampled, the effective characteristic time-scales of the various tracers will decrease, most strongly affecting H${\alpha}$. In this section, we show how incomplete IMF sampling affects the inferred SFR tracer time-scales.

	\subsection{Method for finding the characteristic time-scales for a stochastically sampled IMF}\label{sec:IMFsamplingMethod}
	We adapt the method we present in \cref{sec:methodNonStoc} to investigate the effects of a stochastically sampled IMF. In \cref{sec:IMFsamplingTheory}, we derive how we expect the characteristic time-scales to change as a result of incomplete IMF sampling using a purely analytical approach. Specifically, we consider the probability of finding stars within the star-forming region that are sufficiently massive to generate the desired SFR tracer emission. In this section, we test experimentally if we recover the same behaviour as described there.
	
	We create the reference maps in the same way as before: the reference maps are mass surface density maps of the star particles within the age bins specified in \cref{tab:resultsSolar}.
	However, for the SFR tracer emission maps, we first scale the star particle masses by some mass scaling factor, ${F_{\mathrm{m}}}$, before \textsc{slug2} --- now using it stochastic IMF sampling module --- predicts the expected emission.
	The values of ${F_{\mathrm{m}}}$ range from 0.01--100, where a lower mass scaling factor means the IMF will be less well sampled.
	We then use \textsc{Heisenberg} to determine the characteristic time-scale, as in \cref{sec:resultsSingleTimeScale}.
	The characteristic time-scale we associate to each mass scaling factor is the average of three characteristic time-scales determined from three independently generated stochastic realisations of the synthetic emission maps. This accounts for the spread in time-scales that results from the stochastic nature of the synthetic emission maps.

	We aim to determine the relative change of the SFR tracer time-scale as a function of IMF sampling. To express the latter, we define a characteristic, average star-forming region mass, ${\overline{M}_{\mathrm{r}}}$, as
	\begin{equation}\label{eq:aveRegionMass}
		\overline{M}_{\mathrm{r}} = \Sigma_{\mathrm{SFR}} \times \tau \times \uppi {\left(\frac{\lambda}{2}\right)}^{2}~\mathrm{,}
	\end{equation}
	which uses the SFR surface density, ${\Sigma_{\mathrm{SFR}}}$, and quantities that \textsc{Heisenberg} measures: the total duration of the evolutionary timeline, ${\tau}$, and the typical separation length of independent star-forming regions, ${\lambda}$ \citep[for details see][]{KRUI18}.

	At a fixed total duration of the evolutionary timeline and region separation length, the degree of IMF sampling is controlled by ${\Sigma_{\mathrm{SFR}}}$.
	We calculate the value of ${\Sigma_{\mathrm{SFR}}}$ as
	\begin{equation}\label{eq:SFR}
		\Sigma_{\mathrm{SFR}} = \frac{\sum m_{i}}{t_{\mathrm{E,\,0}}\uppi r^{2}} \times F_{\mathrm{m}}~\mathrm{,}
	\end{equation}
	where ${\sum m_{i}}$ is the total mass of all the star particles that fall within the age bin appropriate for the filter, i.e.\ ${0 \leq \mathrm{Age} \leq t_{\mathrm{E,\,0}}}$ (see \cref{app:completeTable} for the values of ${t_{\mathrm{E,\,0}}}$), which is then scaled by the mass scaling factor ${F_{\mathrm{m}}}$, ${t_{\mathrm{E,\,0}}}$ is the width of that age bin, and ${r}$ is the radius of the galaxy being studied (for our simulated galaxy ${r = 10~\mathrm{kpc}}$, as determined from a visual inspection of the synthetic emission maps).

	In \cref{eq:SFR}, we consider ${\Sigma_{\mathrm{SFR}}}$ as the galaxy average SFR surface density.
	If there are no strong large-scale morphological features, as is the case here, this galaxy average SFR surface density is appropriate to use in the calculation of ${\overline{M}_{\mathrm{r}}}$.
	Otherwise, the expression in \cref{eq:SFR} should account for a non-uniform spatial distribution of star-forming regions across the galaxy by including a factor ${\mathcal{E}_{\mathrm{star,glob}}}$, which indicates the ratio of the mass surface density\footnote{%
		The quantity ${\mathcal{E}_{\mathrm{star,glob}}}$ represents a mass surface density ratio because the reference maps show the mass surface density.
		In typical observational applications, ${\mathcal{E}_{\mathrm{star,glob}}}$ would be a flux density ratio.
	}
	on a size scale of ${\lambda}$ to its area average across the map (see \citealt[Section~3.2.9]{KRUI18} for more details).

	By introducing a \enquote{mass scaling factor}, ${F_{\mathrm{m}}}$, we are able to test experimentally how the characteristic time-scale of different SFR tracers change as a smooth function of IMF sampling.
	We will use these experimental results to see if we observe the behaviour predicted in \cref{sec:IMFsamplingTheory}.

	\subsection{SFR tracer time-scales for a stochastically sampled IMF}\label{sec:resultsStochSampling}
	
	\begin{figure*}
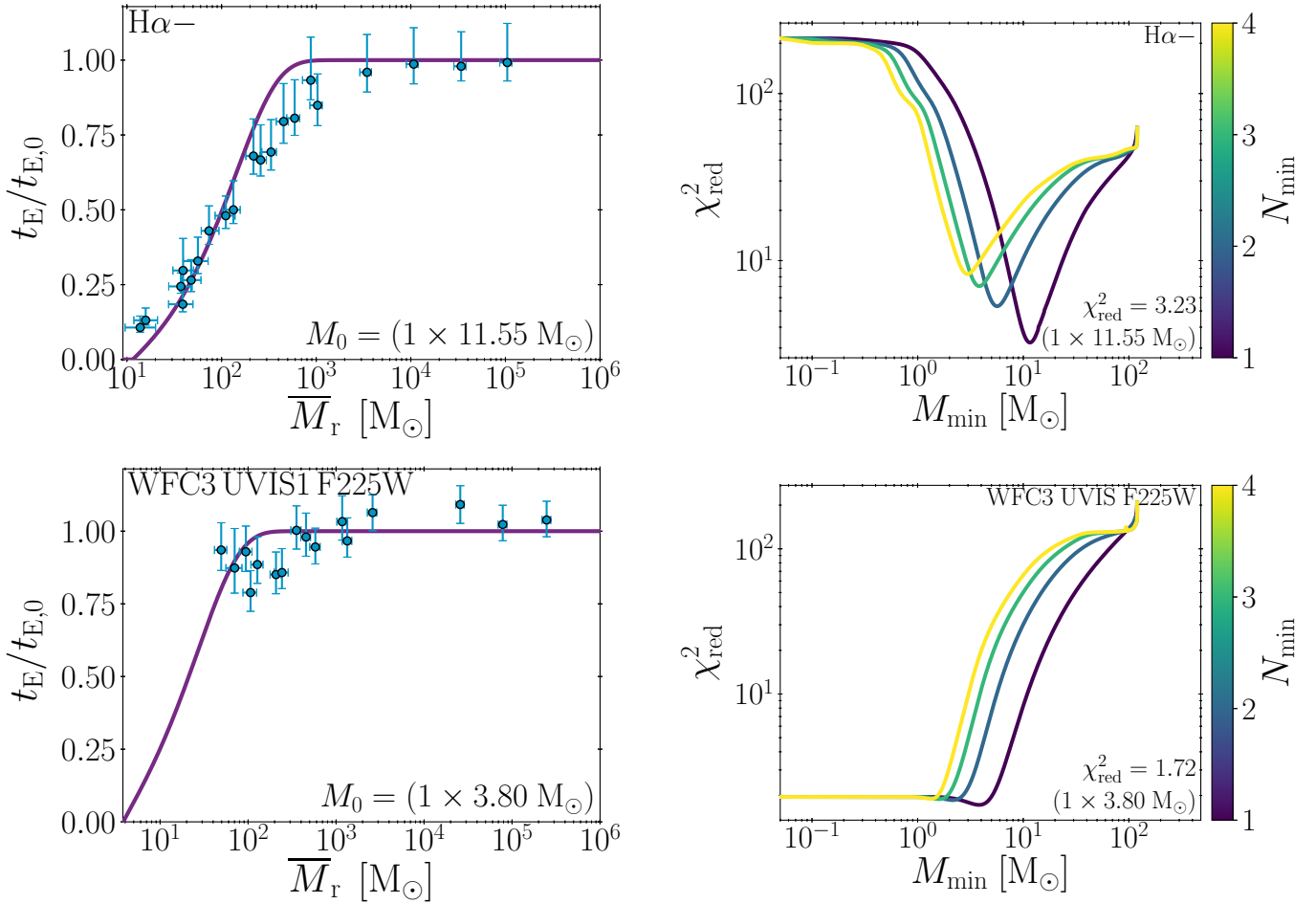

		\centering
		\VerticalCentered{\includegraphics[width=\columnwidth]{\img{modc020-QH0_tgas}}}%
		\hfill%
		\VerticalCentered{\includegraphics[width=\columnwidth]{\img{modc020-QH0_Chi2_Nmin_log}}}%
		\\[\baselineskip]
		\VerticalCentered{\includegraphics[width=\columnwidth]{\img{modc020-WFC3_UVIS_F225W_tgas}}}%
		\hfill%
		\VerticalCentered{\includegraphics[width=\columnwidth]{\img{modc020-WFC3_UVIS_F225W_Chi2_Nmin_log}}}%
		\caption{%
			\textbf{Top row:}~H${\alpha}$ emission excluding the continuum (H${\alpha{-}}$).
			\textbf{Bottom row:}~UV emission (WFC3 UVIS F225W).
			\textbf{Left column:}~Change of the characteristic time-scale of the SFR tracer, relative to the characteristic time-scale we determine from a well sampled IMF, as a function of the average independent star-forming region mass, ${\overline{M}_{\mathrm{r}}}$.
			The data points show the results of the experiments in which we apply \textsc{Heisenberg} to synthetic SFR tracer maps with stochastically sampled IMFs at solar metallicity.
			For comparison, the purple curve shows the best-fitting analytical model from \cref{sec:IMFsamplingTheory}.
			At low region masses, the characteristic time-scales decrease due to the incomplete sampling of the IMF\@.
			\textbf{Right column:}~Change of ${\chi^{2}_{\mathrm{red}}}$ with minimum stellar mass, ${M_{\min}}$, and the minimum number of stars of that mass, ${N_{\min}}$.
			The minimum ${\chi^{2}_{\mathrm{red}}}$ found is indicated in the bottom right with the best-fitting model parameters (${N_{\min} \times M_{\min}}$).
		}\label{fig:results}
	\end{figure*}

	We now present the results of our experiments to test how the characteristic time-scales of H${\alpha}$ and several UV SFR tracers change as a result of incomplete IMF sampling. In \cref{fig:results}, we present the solar-metallicity results for H${\alpha{-}}$ and WFC3 UVIS F225W as examples of how the characteristic time-scales change as a function of the average mass of an independent star-forming region, ${\overline{M}_{\mathrm{r}}}$.
	The quantity ${\overline{M}_{\mathrm{r}}}$ characterises (chiefly through ${\Sigma_{\mathrm{SFR}}}$, see \cref{eq:aveRegionMass}) how well the IMF is sampled: lower values of ${\overline{M}_{\mathrm{r}}}$ result in a more stochastically sampled IMF\@.
	Each data point\footnote{%
		For details on the error calculation on ${\overline{M}_{\mathrm{r}}}$, see \cref{app:ErrorProp}.
	}
    in the two left hand panels of \cref{fig:results} corresponds to a different mass scaling factor, ${F_{\mathrm{m}}}$.
	The quantity shown on the vertical axis, ${t_{\mathrm{E}}/t_{\mathrm{E,\,0}}}$, is the factor by which the measured characteristic time-scale is reduced relative to the time-scale obtained for a well-sampled IMF (as listed in \cref{app:completeTable}), as a result of incomplete IMF sampling at small region masses or low SFR surface densities.
	
	We describe the relation between reduction factor, ${t_{\mathrm{E}}/t_{\mathrm{E,\,0}}}$, and ${\overline{M}_{\mathrm{r}}}$ through the probability ${P\!\left( N \geq N_{\min} \right)}$ as derived in \cref{sec:IMFsamplingTheory}. The purple curves in left-hand panels of \cref{fig:results} indicate the best-fitting form of this function, obtained by varying the minimum stellar mass contributing to the SFR tracer emission ($M_{\min}$) and the number of such stars required ($N_{\min}$).
	We constrain the values for these two free parameters using a brute-force approach: we calculate the value of ${\chi^{2}_{\mathrm{red}}}$ (accounting for the uncertainties on both the abscissa and the ordinate, see \citealt{OREA82}) for a range of ${N_{\min}}=1{-}4$ and ${M_{\min}}=0{-}120~\mathrm{M}_\odot$, and use the minimum ${\chi^{2}_{\mathrm{red}}}$ to define the best-fitting parameter values.
	The right-hand panels of \cref{fig:results} show the dependence of ${\chi^{2}_{\mathrm{red}}}$ on ${N_{\min}}$ and ${M_{\min}}$ for the two example filters. The best fits are always obtained for $N_{\min}=1$.

	When fitting for the two free parameters, we reject data points that exceed the time-scale obtained for a well sampled IMF (i.e.\ ${t_{\mathrm{E}}/t_{\mathrm{E,\,0}}>1}$) by more than ${1\sigma}$.
    This is to remedy an issue at low mass scaling factors (typically ${F_{\mathrm{m}} \approx 0.01}$), where we find that the emission from the continuum can dominate over the SFR tracer when using the H$\alpha{+}$ filter.
	This results in characteristic time-scales that describe the long-lived continuum emission and therefore can be orders of magnitude higher than ${t_{E,\,0}}$.
    The above data selection criterion also affects the UV filters at low mass scaling factors (${F_{\mathrm{m}} \approx 0.01}$), where the emission becomes dominated by UV-faint, low-mass stars.
	As a result, the turn off from ${t_{\mathrm{E}}/t_{\mathrm{E,\,0}} = 1}$ is generally poorly sampled (see \cref{fig:results} for an example), which means that we cannot reliably distinguish between different ${N_{\min}}$ and ${M_{\min}}$. Therefore, we conclude that UV emission is not significantly affected by IMF sampling.

	\begin{table*}
		\centering
		\caption{%
			The functional form of the conversion factor, ${P\left(N \geq N_{\min}\right)}$, between the characteristic time-scale measured for a well-sampled IMF and a stochastically sampled IMF has two parameters, ${N_{\min}}$ and ${M_{\min}}$.
			We use ${N_{\min} = 1}$ and show here the values of ${M_{\min}}$.
		}\label{tab:ZMmin}
		\begin{tabular}{lrrrrr}
			\toprule
			{} & 0.05~${\mathrm{Z_{\odot}}}$ & 0.20~${\mathrm{Z_{\odot}}}$ & 0.40~${\mathrm{Z_{\odot}}}$ & 1.00~${\mathrm{Z_{\odot}}}$ & 2.00~${\mathrm{Z_{\odot}}}$\tabularnewline
			\midrule
			H${\alpha{-}}$ & 11.50 & 11.95 & 13.00 & 11.55 & 10.45\tabularnewline
			H${\alpha{+}}$\,10\,\AA{} & 10.75 & 10.05 & 12.25 & 13.90 & 12.00\tabularnewline
			H${\alpha{+}}$\,20\,\AA{} & 12.10 & 12.05 & 12.40 & 10.35 & 9.95\tabularnewline
			H${\alpha{+}}$\,40\,\AA{} & 9.65 & 9.45 & 7.95 & 8.35 & 10.45\tabularnewline
			H${\alpha{+}}$\,80\,\AA{} & 9.20 & 8.70 & 8.85 & 6.25 & 8.00\tabularnewline
			H${\alpha{+}}$\,160\,\AA{} & 10.20 & 5.20 & 7.80 & 8.60 & 9.35\tabularnewline
			\bottomrule
		\end{tabular}
	\end{table*}

	\cref{tab:ZMmin} lists the best-fitting values of ${M_{\min}}$ (${N_{\min} = 1}$ in all cases) for the full range of metallicities (${Z/\mathrm{Z_{\odot}} = 0.05,~0.20,~0.40,~1.00,~2.00}$) for the H${\alpha}$ filters.
    We find no unambiguous relation between ${M_{\min}}$ and metallicity or filter width. However, smaller filter widths generally have higher ${M_{\min}}$.
	Higher values of ${M_{\min}}$ imply higher star-forming region masses below which IMF sampling affects the SFR tracer time-scale.
	
	We combine the best-fitting values in \cref{tab:ZMmin} with the expressions for ${P\!\left( N \geq N_{\min} \right)}$ provided in \cref{sec:IMFsamplingTheory} to predict the region masses below which incomplete IMF sampling affects the SFR tracer time-scales.
	For ${\mathrm{H\alpha{-}}\left(+\right)}$, this range is ${\overline{M}_{\mathrm{r}} \gtrsim \text{600--800}~\left(\text{200--900}\right)~\mathrm{M_{\odot}}}$.
	For a region separation length of ${\lambda=200}$~pc and a total timeline duration of ${\tau=20}$~Myr \citep[typical for nearby star-forming galaxies, see][]{CHEV20}, these characteristic region mass limits correspond to ${\Sigma_{\mathrm{SFR}} \gtrsim \left(\text{1.0--1.3}\right) \times10^{-3}~\mathrm{M_{\odot}~yr^{-1}~kpc^{-2}}}$ for H${\alpha{-}}$ and ${\Sigma_{\mathrm{SFR}} \gtrsim \left(\text{0.3--1.4}\right) \times10^{-3}~\mathrm{M_{\odot}~yr^{-1}~kpc^{-2}}}$ for H${\alpha{+}}$.

	\cref{fig:results} demonstrates that it is important to consider the effects of IMF sampling at low SFR surface densities, when constraining the characteristic time-scale for the ${\mathrm{H\alpha{\pm}}}$ filters.
	This is because at low SFR surface densities, the massive stars required to produce H${\alpha}$ emission are not always present.
	If we ignore this fact, the ${\mathrm{H\alpha{\pm}}}$ characteristic time-scales will be overestimated; as a result, the evolutionary time-line would be incorrectly calibrated and the time-scales obtained with \textsc{Heisenberg} would also be overestimated.
	The agreement between the results of these experiments and the theoretical model also demonstrate that the IMF sampling theory presented in \cref{sec:IMFsamplingTheory} accurately describes how the characteristic time-scale of ${\mathrm{H\alpha{\pm}}}$ changes due to incomplete IMF sampling.
	This means that observational applications of the \KLPrin{} can use the expressions provided in \cref{eq:emissionTime,eq:Mrmin,eq:fm,eq:ApproxP,eq:aveRegionMass} to derive an SFR tracer time-scale corrected for IMF sampling.
	For the UV tracers, however, we find that the characteristic time-scales are mostly insensitive to the effects of incomplete IMF sampling.

	\section{Comparison to observations}\label{sec:obs}
	This paper predicts the characteristic emission time-scales for SFR tracers, this can only be done using a galaxy simulation because it requires a reference map of which the duration of emission is known exactly.
	For the simulation, we do this by constructing an artificial reference map occupied by the star particles within a known age range and therefore known reference time-scale.
	It is not possible to do this for observed galaxies.
	However, it is possible to test whether the observed \textit{ratio} between the emission time-scales of two different SFR tracers is consistent with our predictions.
	In \citet{KRUI19}, we used our predicted H${\alpha}$ reference time-scale at the half-solar metallicity of NGC300 (${t_{\mathrm{E,\,0}}^{\mathrm{H\alpha}} = 4.59 \pm 0.14}$~Myr) to measure a CO cloud lifetime of ${10.8^{+2.1}_{-1.7}}$~Myr.
	We can now use the measured CO cloud lifetime as a reference time-scale in an experiment combining the CO map (now acting as the reference map) with a UV emission map.
	This allows us to test if the resulting UV emission time-scale is consistent with our prediction for the UV reference time-scale.

\begin{figure}
		\centering
		\includegraphics[width=\columnwidth]{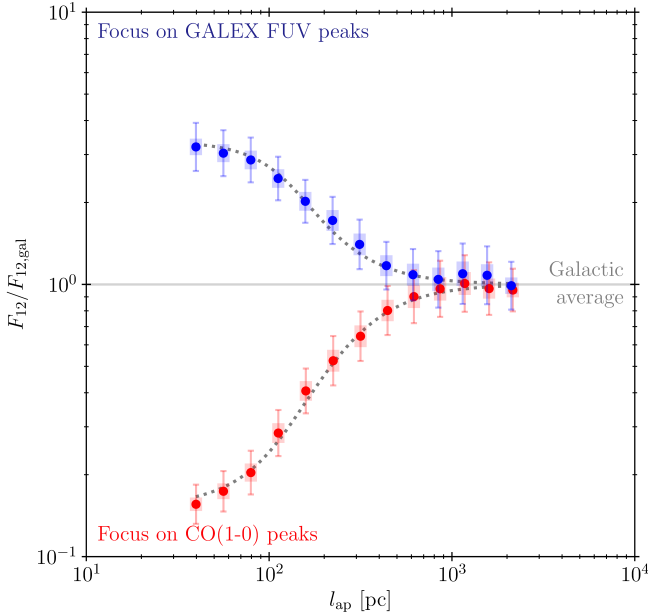}%
		\caption{%
			Tuning fork diagram obtained for the combination of CO(1--0) (Phase 2) and GALEX FUV (Phase 1) emission of NGC300 \citep[see][]{KRUI19}.
			The symbols show the relation between the phase-1-to-phase-2 flux ratio (${F_{12}}$) calculated at the locations of the emission peaks, relative to the galactic-scale phase-1-to-phase-2 flux ratio (${F_{\mathrm{12,\,gal}}}$) as a function of aperture size, ${l_{\mathrm{ap}}}$.
            The error bars indicate the ${1\sigma}$ uncertainty on each individual data point, whereas the shaded areas indicate the effective ${1\sigma}$ uncertainty range that accounts for the covariance between the data points.
            The best-fitting model is indicated by the two dotted curves.
		}\label{fig:ngc300}
	\end{figure}

	As a first test of the accuracy of our inferred time-scales, we combine the GALEX FUV map of NGC300 with the CO data presented in \citet{KRUI19}.
	For the CO map, we use the identical experiment setup as in \citet{KRUI19}, adopting the same set of emission peaks identified there and removing diffuse emission in the same way.
	For the FUV map, the emission peaks on which the apertures are placed are identified over a flux range of ${1.3}$~dex below the brightest peak in the map, using flux contours at intervals of ${0.35}$~dex to separate adjacent peaks.
	In addition, we remove the DC offset from the FUV map by filtering it with a high-pass Gaussian filter in Fourier space on a size scale ${>1000\lambda}$ \citep{HYGA19}.
    Other than these details, we apply the default analysis described in \citet{KRUI18, KRUI19}.
	The resulting tuning fork diagram (also see \cref{fig:exampleTuningfork}) is shown in \cref{fig:ngc300}.
	We obtain a good fit, with a UV emission time-scale of ${t^{\mathrm{FUV}}_{\mathrm{obs}} = 23.1^{+5.9}_{-3.5}}$~Myr.
	Given the half-solar metallicity of NGC300, this should be compared to the reference time-scale predicted by \cref{eq:UVtimeMetalFit} for ${Z = 0.5~\mathrm{Z_{\odot}}}$, which is ${t^{\mathrm{UV}}_{\mathrm{E,\,0}} = 19.2 \pm 2.0}$~Myr.

	Fundamentally, this experiment expresses the reference time-scale for GALEX FUV emission in units of the reference time-scale of continuum-subtracted H${\alpha}$ emission (using CO as an intermediate step).
	This is the case because in both the CO-H${\alpha}$ experiment of \citet{KRUI19} and the CO-FUV experiment carried out here, we have only measured the time-scale ratios.
	We should thus compare the observed and predicted ${t^{\mathrm{FUV}}/t^{\mathrm{H\alpha}}}$ ratio.
	We measure ${t^{\mathrm{FUV}}_{\mathrm{obs}}/t^{\mathrm{H\alpha}}_{\mathrm{obs}} = 5.0^{+1.3}_{-0.8}}$, whereas the calibration of this paper predicts ${t^{\mathrm{FUV}}_{\mathrm{E,\,0}}/t^{\mathrm{H\alpha}}_{\mathrm{E,\,0}}=4.2\pm0.5}$.
	These values agree to within the uncertainties (at ${0.9\sigma}$, or ${\sim20}$~per~cent), which acts as a first demonstration that the reference time-scales derived in this work are consistent with observations.

	Our method yields a measurement of the FUV-to-H${\alpha}$ time-scale \textit{ratio}.
	This means that an arbitrary scaling of both the H${\alpha}$ and FUV characteristic time-scales, for either the predicted or observed ratios, would also result in agreement.
	However, the absolute time-scales individually must also still be physical.
	A comparison of the above numbers to other measurements in the literature shows that they fall within the range of expected values.
	For instance, \citet{LERO12} find that a young stellar population has emitted 50~per~cent of its H${\alpha}$ emission after 1.7~Myr, and 95~per~cent after 4.7~Myr.
	Our characteristic H${\alpha}$ time-scale at solar metallicity of 4.3~Myr falls within this range.
	The same applies for our GALEX FUV time-scale of 17.1~Myr, which falls within the time interval at which 50--95~per~cent of the cumulative flux has been emitted (4.8--65~Myr, see \citealt{LERO12}).
	Future work combining H${\alpha}$ and UV observations of nearby galaxies will enable a more comprehensive test of the presented time-scales.

	\section{Conclusions}\label{sec:conclusions}
	We have applied a new statistical method (the \textsc{Heisenberg} code, which uses the \enquote{uncertainty principle for star formation}, \citealt{KRUI14,KRUI18}) to constrain the characteristic emission time-scales of SFR tracers, i.e.\ the durations over which H${\alpha}$ and UV emission emerge from coeval stellar populations, specifically within the \enquote{uncertainty principle} formalism.
	We expect these time-scales to be critical in a variety of future studies.
	Firstly, observational applications of \textsc{Heisenberg} will enable the empirical characterisation of the cloud lifecycle across a wide range of galactic environments, by measuring e.g.\ the molecular cloud lifetime and the time-scale for cloud destruction by feedback.
	However, in order to lead to physically meaningful constraints, these applications require the use of a known \enquote{reference time-scale} for turning the measured relative time-scales into absolute ones.
	This reference time-scale is provided by the SFR tracer time-scales obtained in this work.
	Secondly, the emission time-scales obtained here and their dependence on metallicity and filter properties provide a helpful point of reference for studies of photoionisation feedback and UV heating.

	To obtain the SFR tracer emission time-scales, we generate synthetic SFR tracer emission maps of a simulated near-${L^{\star}}$ isolated flocculent spiral galaxy using the stochastic SPS code \textsc{slug2} \citep{SILV12, SILV14, KRUM15}.
	We then apply \textsc{Heisenberg} to combinations of these synthetic maps and an independent set of \enquote{reference maps}, which show the star particles from the simulation in specific, known age bins.
	With this approach, we self-consistently measure the characteristic time-scales for H${\alpha}$ emission (with and without continuum subtraction), as well as 12 different UV filters.

	For stellar populations at solar metallicity and with a fully sampled IMF we find the characteristic time-scales for H${\alpha}$ with continuum subtraction to be ${4.3^{+0.1}_{-0.3}}$~Myr, and 5.6--16.4~Myr without.
    For the UV filters, the reference time-scale falls in the range 17.1--33.3~Myr, and nearly monotonically increases with wavelength.
	When considering stellar populations with different metallicities (${Z/\mathrm{Z_{\odot}} = \{0.05,~0.20,~0.40,~1.00,~2.00\}}$) the range of characteristic time-scales increases, to
	3.9--5.5~Myr for H${\alpha}$ with continuum subtraction and 5.1--16.4~Myr without, as well as 14.5--33.3~Myr for the UV filters.
	We define empirical power-law relations that provide the characteristic time-scale as a function of metallicity (\cref{eq:HAtimeMetalFit,eq:HAWidtimeMetalFit,eq:UVtimeMetalFit}).
	These empirical relations include the response-weighted mean wavelength (${\overline{\lambda}_{\mathrm{w}}}$) for UV filters and the filter width (${W}$) for the H${\alpha{+}}$ filters.
	These dependences enable the use of a single expression to determine the characteristic time-scale for all UV and H${\alpha\pm}$ SFR filters for a given combination of filter properties and the metallicity of the environment.

	We also investigate the effects of a stochastically sampled IMF on the characteristic time-scales.
	Incomplete IMF sampling is found to affect the obtained characteristic emission time-scales in low-${\Sigma_{\mathrm{SFR}}}$ galaxies.
	We quantify this dependence by stochastically sampling from the IMF prior to generating the synthetic SFR tracer emission maps and then measuring the characteristic time-scales with \textsc{Heisenberg}.
	We use a \citet{CHAB05} IMF to calculate the probability, ${P}$, of forming at least ${N_{\min}}$ stars of mass ${M_{\min}}$ or higher given a star-forming region mass ${\overline{M}_{\mathrm{r}}}$.
	We then demonstrate that this probability is a good predictor for the ratio between the characteristic time-scale for a stochastically sampled IMF, ${t_{\mathrm{E}}}$, and that of a well-sampled IMF, ${t_{\mathrm{E,\,0}}}$.
	As a result, we obtain a relation between ${t_{\mathrm{E}}/t_{\mathrm{E,\,0}}}$ and the characteristic mass of independent star-forming regions, ${\overline{M}_{\mathrm{r}}}$.
	Given an SFR surface density (from which the characteristic region mass can be derived), this relation quantifies the relative change of the SFR tracer time-scale due to IMF sampling as a function of the galactic environment.

	For UV tracers, the impact of IMF sampling on the characteristic time-scale is minimal (${<30}$~per~cent) and can therefore be ignored (this applies to all metallicities).
	However, incomplete IMF sampling has a significant effect on the characteristic time-scales of H${\alpha}$ emission.
	At low SFR surface densities, the H${\alpha}$ emission time-scale is suppressed due to IMF sampling effects.
	Depending on the metallicity and on whether the continuum emission has been subtracted, the characteristic time-scale for a well sampled IMF can be used for ${\overline{M}_{\mathrm{r}} \gtrsim \text{200--900}~\mathrm{M_{\odot}}}$, which for a region separation length of ${\lambda=200}$~pc and a total timeline duration of ${\tau=30}$~Myr corresponds to ${\Sigma_{\mathrm{SFR}} \gtrsim \left(\text{0.3--1.3}\right) \times10^{-3}~\mathrm{M_{\odot}~yr^{-1}~kpc^{-2}}}$.
    However, at lower region masses or SFR surface densities, the H${\alpha}$ reference time-scale must be corrected to account for the effects of IMF sampling.
	We derive fitting functions describing the change of the H${\alpha}$ time-scales as a function of the average independent star-forming region mass, ${\overline{M}_{\mathrm{r}}}$, as parametrised by the minimum stellar mass required for H${\alpha}$ emission, ${M_{\min}}$, which we tabulate as a function of metallicity (\cref{eq:emissionTime,eq:Mrmin,eq:fm,eq:ApproxP,eq:aveRegionMass,eq:SFR} as well as \cref{tab:ZMmin}).

	Even though we have arrived at the above reference time-scales by carrying out a set of numerical experiments using a galaxy simulation, and one could thus argue that the results are model-dependent, we reiterate that the results are not expected to be sensitive to the details of the baryonic physics in the simulations (see discussion in \cref{sec:methodNonStoc}).
    The measurements carried out in this work require a physically-motivated correlation of positions and ages of star particles.
    The critical goal was to carry out these measurements self-consistently within the framework of our method and thus enabling its future observational applications.
    In principle, this measurement could have been performed using maps of randomly-generated distributions of regions: fundamentally, we have only characterised how quickly young stellar emission fades in the adopted SPS model.
    However, the main advantage of using a galaxy simulation is that it generates a distribution with a physically reasonable imprint of galactic morphology and the positional correlation of star formation events by self-gravity and stellar feedback.
    The accuracy of the results is demonstrated by a first comparison to observations of H${\alpha}$ and GALEX FUV emission in the nearby galaxy NGC300 (\cref{sec:obs}), which shows that the time-scales predicted by this work are consistent with the observed time-scales.

	In summary, we have measured the characteristic emission time-scales of SFR tracers within the \enquote{uncertainty principle} formalism, as a function of metallicity and (for UV and H${\alpha{+}}$) filter properties, as well as their sensitivity to IMF sampling, which effectively expresses their dependence on the SFR surface density.
	This spans the range of key environmental factors that affect the time-scales of H${\alpha}$ and UV emission, and provides important constraints on the duration of photoionisation feedback and UV heating.
	The reference time-scales derived in this work enable observational applications of the \enquote{uncertainty principle for star formation}, in which they are used to turn the relative durations of evolutionary phases into an absolute timeline.
	Specifically, we expect that the fitting functions provided in \cref{eq:HAtimeMetalFit,eq:HAWidtimeMetalFit,eq:UVtimeMetalFit} and \cref{eq:Mrmin,eq:fm,eq:ApproxP} will have great practical use because they enable the straightforward calculation of the reference time-scales as a function of metallicity, UV filter wavelength, and SFR surface density.
	Indeed, the first applications of this method have already used these equations to infer the time-scales driving cloud evolution, star formation, and feedback \citep[as well as \cref{sec:obs} of this paper]{KRUI19, CHEV20b,CHEV20,HYGA19b,WARD20}.
	In view of the variety of recent and upcoming applications of this method, the time-scales presented in this work represent an essential ingredient towards empirically constraining the physics driving molecular cloud lifecycle.

	\section*{Acknowledgements}
	We thank the anonymous referee for a helpful and constructive report that improved the paper.
	The authors acknowledge support by the state of Baden-W{\"u}rttemberg through bwHPC and the German Research Foundation (DFG) through grant INST 35/1134--1 FUGG and INST 37/935--1 FUGG\@.
	DTH and APSH are fellows of the International Max Planck Research School for Astronomy and Cosmic Physics at the University of Heidelberg (IMPRS-HD).
	JMDK gratefully acknowledges funding from the European Research Council (ERC) under the European Union's Horizon 2020 research and innovation programme via the ERC Starting Grant MUSTANG (grant agreement number 714907).
	JMDK and MC gratefully acknowledge funding from the German Research Foundation (DFG) in the form of an Emmy Noether Research Group (grant number KR4801/1--1) and the DFG Sachbeihilfe (grant number KR4801/2--1).
	MRK acknowledges support from the Australia Research Council's Discovery Projects and Future Fellowship funding schemes, awards DP160100695 and FT180100375.
	DTH, JMDK, MC, and MRK acknowledge support from the Australia-Germany Joint Research Cooperation Scheme (UA-DAAD, grant number 57387355).

    \section*{Data availability}
    The data underlying this article will be shared on reasonable request to the corresponding author.



	\bibliographystyle{mnras}
	\bibliography{SFR_Tracer_Lifetime.bib} 


	\appendix

	\section{H\texorpdfstring{${\alpha}$}{α} Absorption and Emission Features}\label{app:HAAbsEmi}
	We produce synthetic emission maps by passing the age and mass information of all the star particles from our simulation to \textsc{slug2} \citep{SILV12, SILV14, KRUM15}.
	\textsc{slug2} then calculates the predicted emission spectrum for each particle, to which we apply UV and H${\alpha{+}}$ filters (H${\alpha{-}}$ comes directly from the hydrogen-ionizing photon emission).
	However, the emission spectrum that \textsc{slug2} produces does not include the underlying H${\alpha}$ absorption from the stellar continuum.
	In this appendix, we use \textsc{Starburst99} \citep{LEIT99, VAZQ05} to investigate when the H${\alpha}$ absorption feature can no longer be neglected.

	We ran \textsc{Starburst99} for an instantaneous burst of star formation for the five standard Geneva evolutionary tracks using a \citet{KROU01} IMF and output the data in 0.1~Myr time steps for 20~Myr.
	We otherwise used the default settings.

	The equivalent width of the H${\alpha}$ emission is taken directly from the \textsc{Starburst99} output files.
	To determine the equivalent width of the absorption feature, we model the continuum (straight line) and the absorption feature (Voigt profile) of the high resolution spectral data in the wavelength range ${6482~\text{\AA{}} \le \lambda \le 6642~\text{\AA{}}}$.

\begin{figure*}
		\centering
		\includegraphics[width=\textwidth]{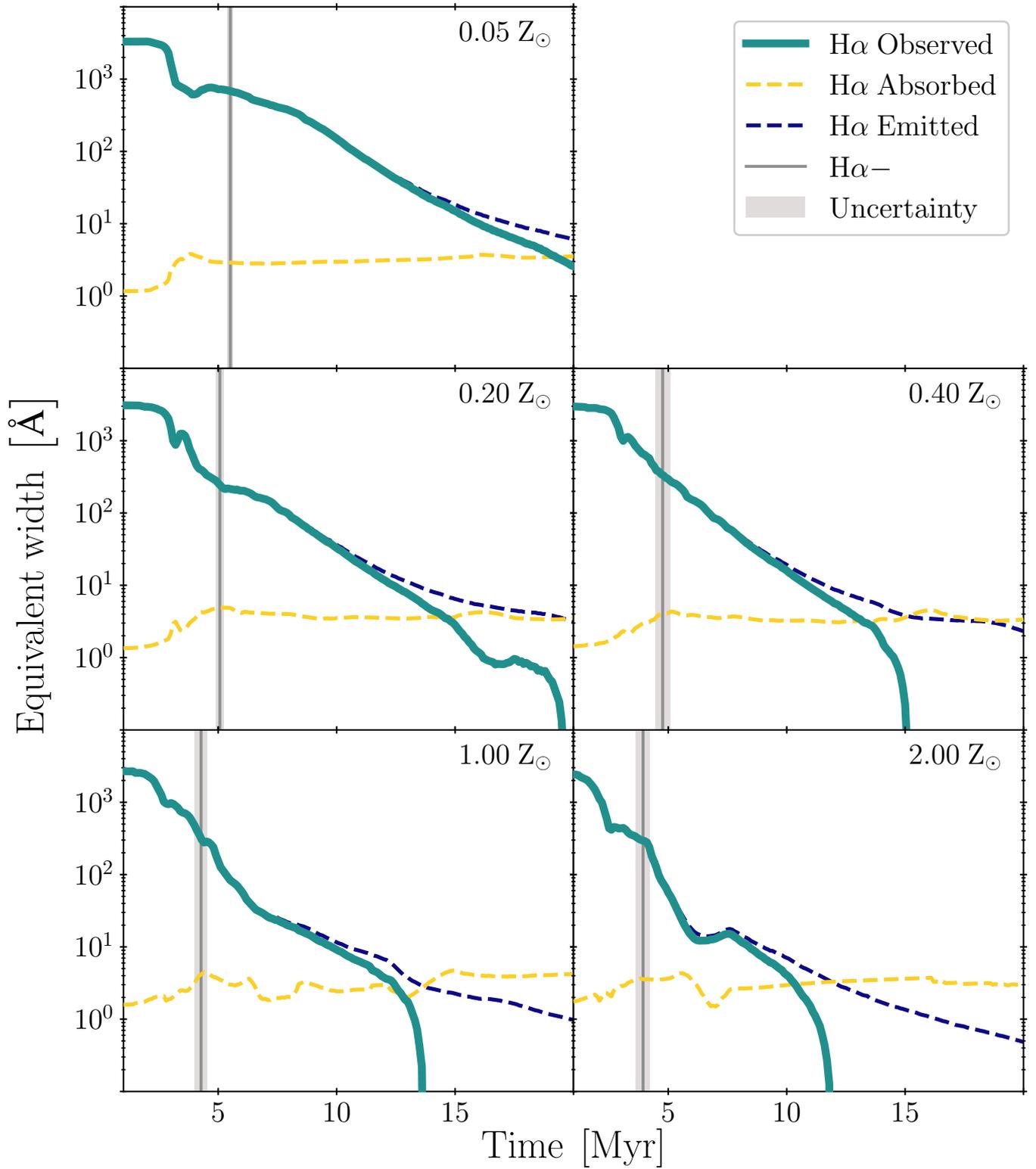}%
		\caption{%
			The results of \textsc{Starburst99} simulations for an instantaneous burst of star formation at 0~Myr.
			We show the change in equivalent width of the H${\alpha}$ absorption and emission feature; we also include the difference between the two equivalent widths (H${\alpha}$ Observed).
			The H${\alpha{-}}$ time-scale is marked for comparison.
		}\label{fig:HAAbsEmi}
	\end{figure*}

	In \cref{fig:HAAbsEmi}, we show the change in the equivalent width of the absorption and emission feature over time; the change in the difference between the two equivalent widths is also included.
	We see that the emission feature is dominant up to at least 10~Myr and longer for the lower metallicities; this is at least 5~Myr longer than the H${\alpha{-}}$ time-scales we measure (see \cref{tab:resultsAllZ}) which are also marked in \cref{fig:HAAbsEmi}.

	We can see from \cref{fig:HAAbsEmi} that the H${\alpha}$ time-scales we are considering fall comfortably within the emission-dominant regime and conclude that the absorption feature can safely be neglected for our analysis.

	\section{Influence of the weighting scheme on the results} \label{app:weights}
	
	\begin{figure}
		\centering
		\includegraphics[width=\columnwidth]{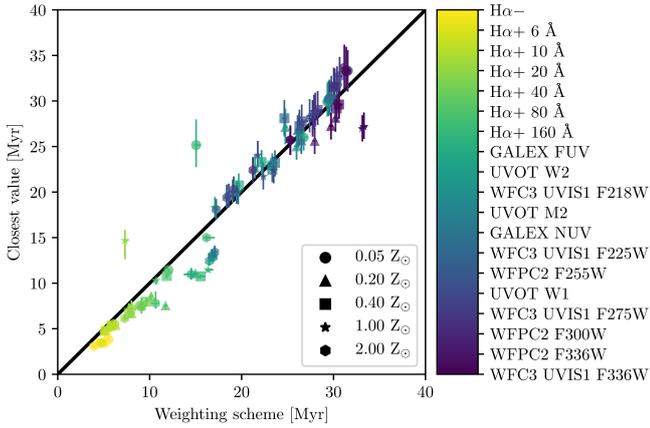}%
		\caption{%
			Comparison of the time-scales measured using the weighting scheme adopted in this paper ($x$-axis) and the time-scales that would have been obtained by taking the value of $t_{\mathrm{E,\,0}}$ that is closest to $t_{\mathrm{M}}$ and $t_{\mathrm{R}}$ ($y$-axis). The data points are colour-coded by filter (see the colour bar), with the shape denoting the metallicity. This figure shows that both methods are in rough agreement, but that there exist clear outliers. For these cases, i.e.\ when both methods disagree, we demonstrate in the text and in \cref{fig:table_weights} that the weighting scheme provides the most accurate result.
		}\label{fig:plot_W}
	\end{figure}
	
	\begin{figure*}
		\centering
		\includegraphics[width=\textwidth]{\img{modc020_H_ALPHA_20_weights}}%
		\caption{%
			Example showing the same table as in \cref{fig:ageBinTimeScales}, but this time for H$\alpha$ including the continuum (H${\alpha{+}}$) with a 20~\AA\ filter. As before, the reference maps are characterised by the age bin used to select the star particles which are included in the reference map. ${t_{\mathrm{M}}}$ denotes the minimum age of the star particles and ${t_{\mathrm{R}}}$ the width of the age bin. The colour-coding is based on the weighting, ${\mathcal{W}}$, used when calculating the weighted average. All values within the tables are given in Myr. This figure shows that the simple approach of taking the measured value that is closest to $t_M$ and $t_R$ is sensitive to local minima (which would in this case overestimate the time-scale by a factor of two, see the offset star in \cref{fig:plot_W}), whereas the weighting scheme adopted in this work avoids this problem by calculating a weighted average of all cells (see the text for details).
		}\label{fig:table_weights}
	\end{figure*}
	
	In \cref{sec:resultsSingleTimeScale}, we adopt a simple weighting scheme to combine the measured emission time-scales for different choices of $t_M$ and $t_R$ (see \cref{fig:ageBinTimeScales}) into a single value. Because our method performs best when $t_{\mathrm{E,\,0}}\sim t_{\mathrm{M}}\sim t_{\mathrm{R}}$ \citep{KRUI18}, this scheme calculates the geometric distance in logarithmic space between the measured time-scale and $t_M$ and $t_R$, and then defines the weight of each cell as its inverse ($\mathcal{W}^{\mathrm{d}}_{ij}$, \cref{eq:distance}). As is common practice, we also weight each cell by the inverse square of the uncertainty ($\mathcal{W}^{\mathrm{u}}_{ij}$, \cref{eq:uncertainty}). A much simpler approach would be to simply adopt the cell for which the measured time-scale is closest to $t_M$ and $t_R$. In this appendix, we briefly demonstrate that this simple approach generally provides similar results to the adopted weighting scheme, but is more sensitive to stochasticity and outliers. This motivates the use of the adopted weighting scheme.
	
	In \cref{fig:plot_W}, we show the comparison of the emission time-scales obtained with both of the above approaches. As the figure shows, both approaches agree to within $\la20$~per~cent, which largely validates our approach. However, there are also clear outliers, with deviations up to a factor of 2. To investigate what causes these outliers, we show the full table of measurements for one of them (H$\alpha$+ 20~\AA) in \cref{fig:table_weights}. This figure shows a fragmented weighting landscape, with two local minima around $t_{\mathrm{E,\,0}}=6{-}7$~Myr and $t_{\mathrm{E,\,0}}\approx15$~Myr. Due to the extreme proximity of the second of these two minima to the corresponding $t_M$ and $t_R$, it has a high weight, even thought the first minimum is the correct answer. This is illustrated by (1) the fact that the surrounding cells have high weights too (i.e.\ it is not isolated) and (2) the fact that the first minimum is more consistent with the measurements for other metallicities and filter widths listed in \cref{app:completeTable}. Had we simply used the cell for which $t_{\mathrm{E,\,0}}$ is closest to $t_M$ and $t_R$, we would have selected the second minimum. By contrast, the adopted weighting scheme ends up with $t_{\mathrm{E,\,0}}=7.3^{+0.4}_{-0.2}$~Myr. This is very close to the correct answer found near the first minimum and shows that the second minimum has a minor effect on the final result when using the adopted weighting scheme.
	
	In principle, we could have adopted a different functional form for \cref{eq:distance}. However, this would necessarily have been more ad hoc than a simple inverse dependence on the geometric distance in logarithmic space used here. We therefore prefer to adopt the simplest approach that avoids a strong sensitivity to outliers, which we have done in \cref{sec:resultsSingleTimeScale}.

    \section{Comparison of the analytical time-scale fits to the measurements} \label{app:fits}
    
    \begin{figure*}
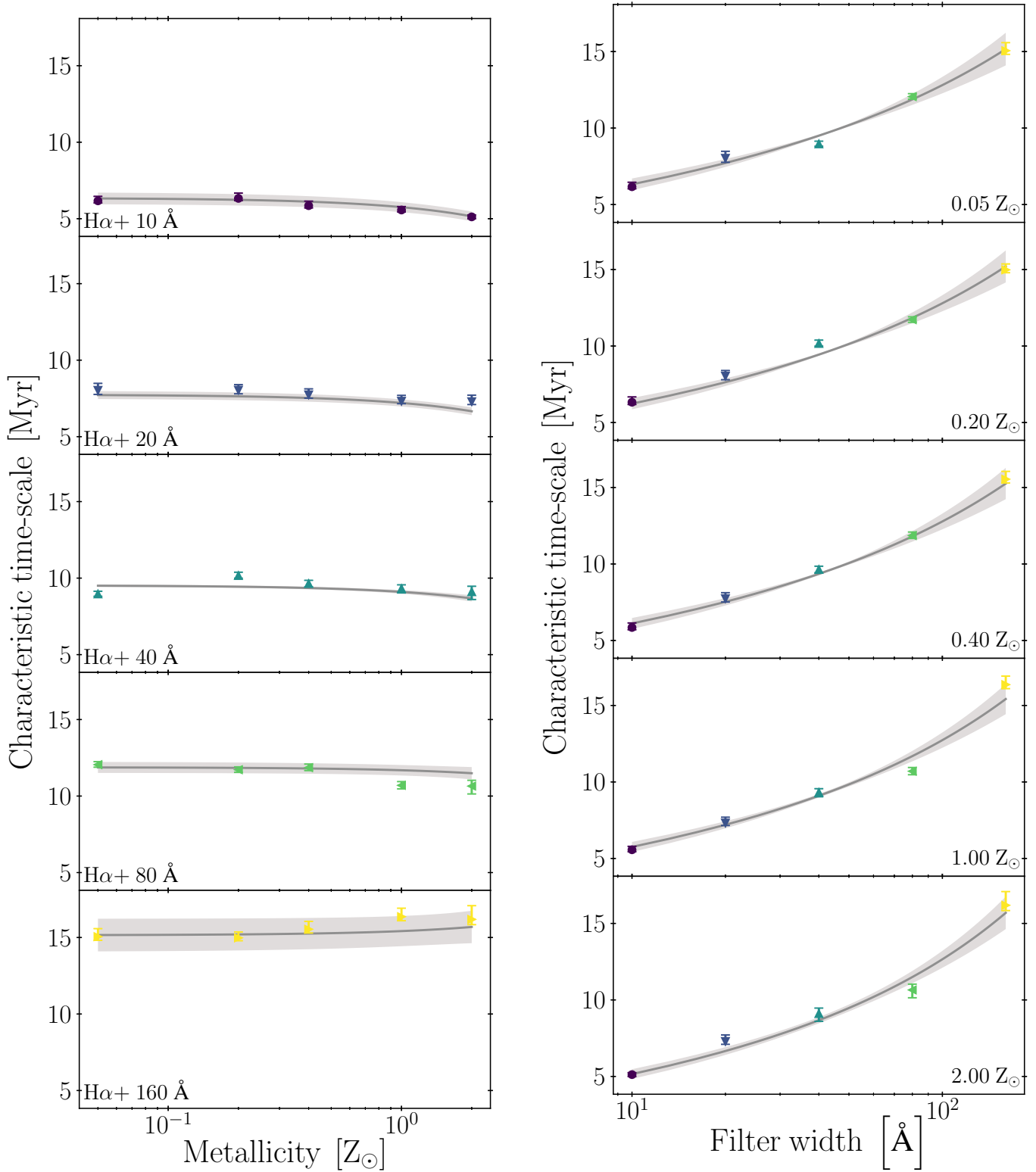

		\centering
		\includegraphics[width=\columnwidth]{\img{ZWid-slice_width}}%
		\hfill%
		\includegraphics[width=\columnwidth]{\img{ZWid-slice_metallicity}}%
		\caption{%
			Characteristic time-scales of H${\alpha{+}}$ filters as a function of metallicity (left) and filter width (right).
			The symbols show the results of applying the \textsc{Heisenberg} code to synthetic H${\alpha{+}}$ maps.
			The grey curve shows the fit from \cref{eq:HAWidtimeMetalFit} and the shaded region indicates the associated uncertainty.
			The symbols and colours in the right-hand panels correspond to those used on the left.
		}\label{fig:ZHWidthfits}
	\end{figure*}

	\begin{figure*}
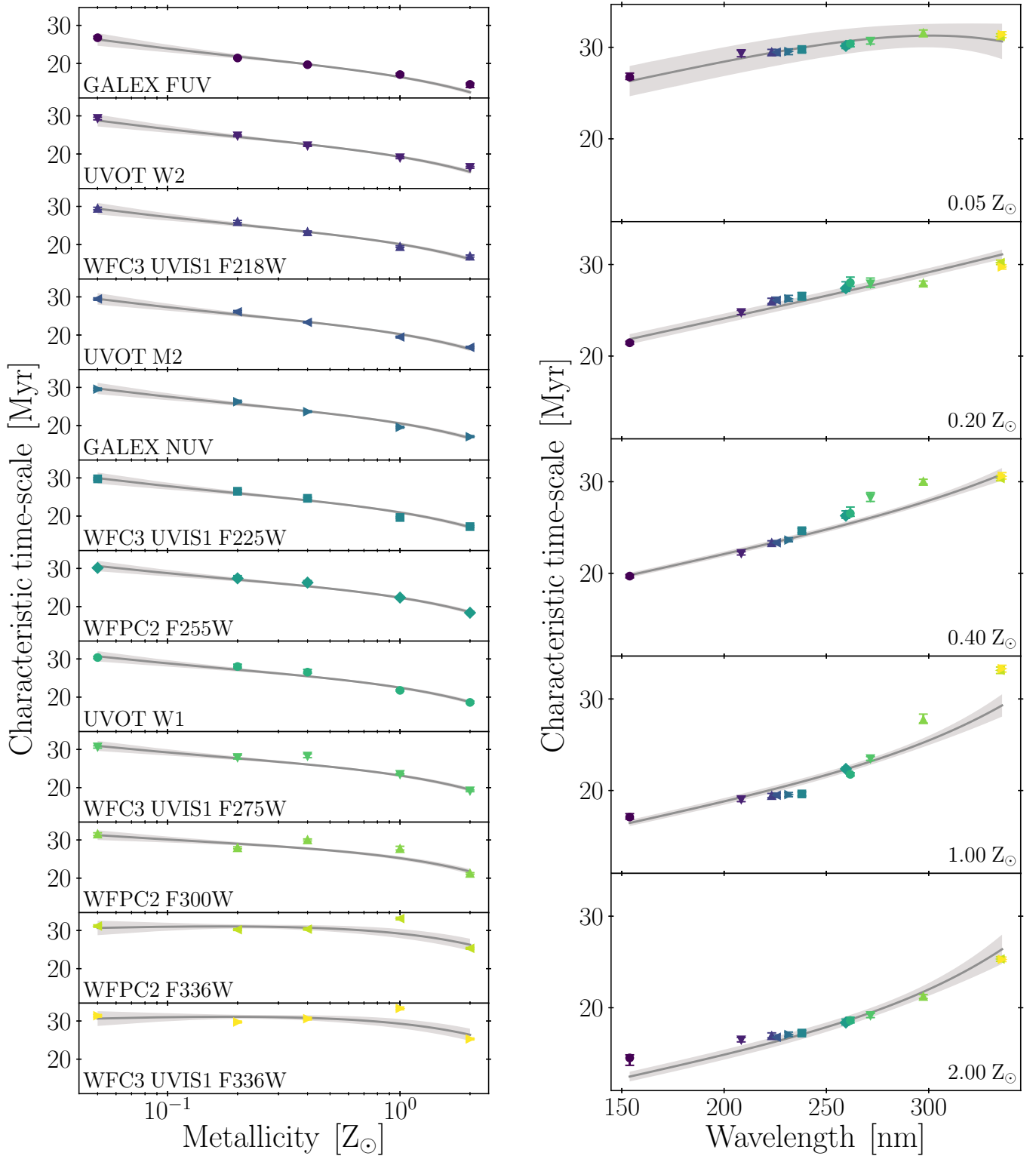

		\centering
		\includegraphics[width=\columnwidth]{\img{ZWav-slice_wavelength}}%
		\hfill%
		\includegraphics[width=\columnwidth]{\img{ZWav-slice_metallicity}}%
		\caption{%
			Characteristic time-scales of UV SFR tracers as a function of metallicity (left) and wavelength (right).
			The symbols show the results of applying the \textsc{Heisenberg} code to synthetic UV maps.
			The grey curve shows the fit from \cref{eq:UVtimeMetalFit} and the shaded region indicates the associated uncertainty.
			The symbols and colours in the right-hand panels correspond to those used on the left.
		}\label{fig:ZWfits}
	\end{figure*}
	
    In \cref{fig:ZHWidthfits} and \cref{fig:ZWfits}, we compare the best-fitting analytical functions for predicting the H$\alpha{+}$ and UV SFR tracer time-scales (Equations~\ref{eq:HAWidtimeMetalFit} and~\ref{eq:UVtimeMetalFit}) to the measurements obtained with \textsc{Heisenberg}. These one-dimensional projections of \cref{fig:ZWid_surface,fig:ZW_surface} are useful for identifying parts of parameter space in which the analytical fits do not describe the measurements well, i.e.\ at long UV wavelengths and solar metallicity (which is the largest discrepancy found across the metallicity-wavelength space considered). In such cases, it might be advisable to use the measurement rather than the analytical expression.

	\section{Complete set of SFR tracer emission time-scales}\label{app:completeTable}

	\begin{table*}
		\caption{%
			A summary of all the characteristic time-scales and corresponding age bins (for producing reference maps in \cref{sec:Metallicity,sec:IMFsampling}), for the different star formation rate tracers (see \cref{tab:filterList} for details).
			These results are for a well sampled IMF\@.
			The filter order is in increasing filter width (${W}$) for H${\alpha{+}}$ and increasing response-weighted mean wavelength (${\overline{\lambda}_{\mathrm{w}}}$) for UV\@.
			\cref{tab:resultsSolar} is included in these tables under the 1.00~${\mathrm{Z_{\odot}}}$ heading.
		}\label{tab:resultsAllZ}
		\subfloat[Characteristic time-scales, ${t_{\mathrm{E,\,0}}}$.]{%
			\begin{minipage}{\textwidth}
				\centering
				\begin{tabular}{lrrrrrrrrrr}
					\toprule
					{} & 0.05~${\mathrm{Z_{\odot}}}$ & 0.20~${\mathrm{Z_{\odot}}}$ & 0.40~${\mathrm{Z_{\odot}}}$ & 1.00~${\mathrm{Z_{\odot}}}$ & 2.00~${\mathrm{Z_{\odot}}}$\tabularnewline
					\midrule
					H${\alpha{-}}$ & ${5.5^{+0.2}_{-0.1}}$ & ${5.1^{+0.1}_{-0.2}}$ & ${4.8^{+0.1}_{-0.3}}$ & ${4.3^{+0.1}_{-0.3}}$ & ${3.9^{+0.1}_{-0.3}}$\tabularnewline
					H${\alpha{+}}$\,10\,\AA{} & ${6.2^{+0.3}_{-0.1}}$ & ${6.3^{+0.3}_{-0.1}}$ & ${5.9^{+0.3}_{-0.1}}$ & ${5.6^{+0.2}_{-0.1}}$ & ${5.1^{+0.1}_{-0.1}}$\tabularnewline
					H${\alpha{+}}$\,20\,\AA{} & ${8.0^{+0.5}_{-0.3}}$ & ${8.0^{+0.4}_{-0.2}}$ & ${7.7^{+0.4}_{-0.2}}$ & ${7.3^{+0.4}_{-0.2}}$ & ${7.3^{+0.4}_{-0.2}}$\tabularnewline
					H${\alpha{+}}$\,40\,\AA{} & ${9.0^{+0.2}_{-0.2}}$ & ${10.2^{+0.2}_{-0.2}}$ & ${9.6^{+0.2}_{-0.2}}$ & ${9.3^{+0.2}_{-0.3}}$ & ${9.1^{+0.4}_{-0.5}}$\tabularnewline
					H${\alpha{+}}$\,80\,\AA{} & ${12.1^{+0.2}_{-0.2}}$ & ${11.7^{+0.2}_{-0.2}}$ & ${11.9^{+0.2}_{-0.2}}$ & ${10.7^{+0.2}_{-0.2}}$ & ${10.7^{+0.4}_{-0.5}}$\tabularnewline
					H${\alpha{+}}$\,160\,\AA{} & ${15.1^{+0.5}_{-0.2}}$ & ${15.0^{+0.4}_{-0.2}}$ & ${15.5^{+0.5}_{-0.2}}$ & ${16.4^{+0.6}_{-0.3}}$ & ${16.2^{+0.9}_{-0.3}}$\tabularnewline
					GALEX FUV & ${26.7^{+0.4}_{-0.3}}$ & ${21.4^{+0.2}_{-0.2}}$ & ${19.7^{+0.2}_{-0.2}}$ & ${17.1^{+0.4}_{-0.2}}$ & ${14.5^{+0.3}_{-0.8}}$\tabularnewline
					UVOT W2 & ${29.3^{+0.3}_{-0.4}}$ & ${24.8^{+0.2}_{-0.2}}$ & ${22.2^{+0.2}_{-0.2}}$ & ${19.0^{+0.3}_{-0.2}}$ & ${16.5^{+0.3}_{-0.2}}$\tabularnewline
					WFC3 UVIS1 F218W & ${29.5^{+0.3}_{-0.4}}$ & ${26.0^{+0.3}_{-0.3}}$ & ${23.3^{+0.2}_{-0.2}}$ & ${19.4^{+0.2}_{-0.2}}$ & ${16.9^{+0.3}_{-0.2}}$\tabularnewline
					UVOT M2 & ${29.5^{+0.3}_{-0.4}}$ & ${26.1^{+0.3}_{-0.3}}$ & ${23.3^{+0.2}_{-0.2}}$ & ${19.5^{+0.2}_{-0.2}}$ & ${16.8^{+0.3}_{-0.2}}$\tabularnewline
					GALEX NUV & ${29.5^{+0.3}_{-0.3}}$ & ${26.3^{+0.4}_{-0.3}}$ & ${23.6^{+0.3}_{-0.2}}$ & ${19.6^{+0.2}_{-0.2}}$ & ${17.1^{+0.3}_{-0.2}}$\tabularnewline
					WFC3 UVIS1 F225W & ${29.8^{+0.2}_{-0.3}}$ & ${26.5^{+0.4}_{-0.3}}$ & ${24.7^{+0.3}_{-0.2}}$ & ${19.6^{+0.2}_{-0.2}}$ & ${17.2^{+0.3}_{-0.2}}$\tabularnewline
					WFPC2 F255W & ${30.1^{+0.3}_{-0.3}}$ & ${27.4^{+0.7}_{-0.3}}$ & ${26.3^{+0.5}_{-0.3}}$ & ${22.4^{+0.2}_{-0.2}}$ & ${18.4^{+0.4}_{-0.3}}$\tabularnewline
					UVOT W1 & ${30.4^{+0.3}_{-0.3}}$ & ${28.0^{+0.6}_{-0.4}}$ & ${26.5^{+0.7}_{-0.3}}$ & ${21.8^{+0.2}_{-0.2}}$ & ${18.6^{+0.4}_{-0.3}}$\tabularnewline
					WFC3 UVIS1 F275W & ${30.7^{+0.3}_{-0.3}}$ & ${27.9^{+0.6}_{-0.4}}$ & ${28.3^{+0.6}_{-0.4}}$ & ${23.5^{+0.2}_{-0.2}}$ & ${19.1^{+0.2}_{-0.2}}$\tabularnewline
					WFPC2 F300W & ${31.6^{+0.3}_{-0.4}}$ & ${28.0^{+0.2}_{-0.3}}$ & ${30.0^{+0.2}_{-0.3}}$ & ${27.7^{+0.6}_{-0.3}}$ & ${21.2^{+0.2}_{-0.2}}$\tabularnewline
					WFPC2 F336W & ${31.2^{+0.3}_{-0.3}}$ & ${30.2^{+0.3}_{-0.2}}$ & ${30.3^{+0.3}_{-0.2}}$ & ${33.1^{+0.4}_{-0.3}}$ & ${25.3^{+0.3}_{-0.2}}$\tabularnewline
					WFC3 UVIS1 F336W & ${31.4^{+0.3}_{-0.3}}$ & ${29.7^{+0.2}_{-0.2}}$ & ${30.6^{+0.4}_{-0.3}}$ & ${33.3^{+0.4}_{-0.4}}$ & ${25.3^{+0.2}_{-0.2}}$\tabularnewline
					\bottomrule
				\end{tabular}
			\end{minipage}
		}\\%
		\subfloat[Age bins, ${t_{\mathrm{E,\,0}} \le \mathrm{Age} \le 2t_{\mathrm{E,\,0}}}$.]{%
			\begin{minipage}{\textwidth}
				\centering
				\begin{tabular}{lrrrrrrrrrr}
					\toprule
					{} & 0.05~${\mathrm{Z_{\odot}}}$ & 0.20~${\mathrm{Z_{\odot}}}$ & 0.40~${\mathrm{Z_{\odot}}}$ & 1.00~${\mathrm{Z_{\odot}}}$ & 2.00~${\mathrm{Z_{\odot}}}$\tabularnewline
					\midrule
					H${\alpha{-}}$ & 5.5--11.0 & 5.1--10.1 & 4.8--9.5 & 4.3--8.6 & 3.9--7.9\tabularnewline
					H${\alpha{+}}$ 10 \AA{} & 6.2--12.3 & 6.3--12.7 & 5.9--11.7 & 5.6--11.1 & 5.1--10.2\tabularnewline
					H${\alpha{+}}$ 20 \AA{} & 8.0--16.1 & 8.0--16.1 & 7.7--15.5 & 7.3--14.6 & 7.3--14.6\tabularnewline
					H${\alpha{+}}$ 40 \AA{} & 9.0--17.9 & 10.2--20.3 & 9.6--19.3 & 9.3--18.6 & 9.1--18.2\tabularnewline
					H${\alpha{+}}$ 80 \AA{} & 12.1--24.1 & 11.7--23.4 & 11.9--23.7 & 10.7--21.4 & 10.7--21.3\tabularnewline
					H${\alpha{+}}$ 160 \AA{} & 15.1--30.1 & 15.0--30.0 & 15.5--31.1 & 16.4--32.7 & 16.2--32.4\tabularnewline
					GALEX FUV & 26.7--53.5 & 21.4--42.9 & 19.7--39.4 & 17.1--34.2 & 14.5--29.0\tabularnewline
					UVOT W2 & 29.3--58.7 & 24.8--49.5 & 22.2--44.3 & 19.0--38.0 & 16.5--33.0\tabularnewline
					WFC3 UVIS1 F218W & 29.5--59.0 & 26.0--52.0 & 23.3--46.7 & 19.4--38.9 & 16.9--33.9\tabularnewline
					UVOT M2 & 29.5--58.9 & 26.1--52.2 & 23.3--46.7 & 19.5--39.0 & 16.8--33.5\tabularnewline
					GALEX NUV & 29.5--59.1 & 26.3--52.5 & 23.6--47.3 & 19.6--39.1 & 17.1--34.1\tabularnewline
					WFC3 UVIS1 F225W & 29.8--59.5 & 26.5--53.0 & 24.7--49.3 & 19.6--39.3 & 17.2--34.5\tabularnewline
					WFPC2 F255W & 30.1--60.3 & 27.4--54.8 & 26.3--52.6 & 22.4--44.7 & 18.4--36.8\tabularnewline
					UVOT W1 & 30.4--60.8 & 28.0--56.0 & 26.5--53.1 & 21.8--43.5 & 18.6--37.2\tabularnewline
					WFC3 UVIS1 F275W & 30.7--61.3 & 27.9--55.8 & 28.3--56.5 & 23.5--47.0 & 19.1--38.3\tabularnewline
					WFPC2 F300W & 31.6--63.1 & 28.0--55.9 & 30.0--60.0 & 27.7--55.4 & 21.2--42.5\tabularnewline
					WFPC2 F336W & 31.2--62.3 & 30.2--60.4 & 30.3--60.7 & 33.1--66.3 & 25.3--50.6\tabularnewline
					WFC3 UVIS1 F336W & 31.4--62.8 & 29.7--59.5 & 30.6--61.2 & 33.3--66.6 & 25.3--50.6\tabularnewline
					\bottomrule
				\end{tabular}
			\end{minipage}
		}
	\end{table*}
	
	In \cref{tab:resultsAllZ}, we list the complete set of SFR tracer emission time-scales constrained in this paper.
	This contains the characteristic time-scales of ${\mathrm{H\alpha{\pm}}}$ and all 12 UV filters, for the five different metallicities ${Z/\mathrm{Z_{\odot}} = 0.05,~0.20,~0.40,~1.00,~2.00}$.
	In addition, we include the age intervals that we adopt to define the stellar reference maps used when measuring the SFR tracer time-scales with the \textsc{Heisenberg} code.
	For more details on the calculations, see \cref{sec:resultsSingleTimeScale}.

	\section{Theoretical expectations for the effect of incomplete IMF sampling}\label{sec:IMFsamplingTheory}
	Here, we predict the relationship between how well the IMF is sampled and the characteristic time-scale of the SFR tracer.
	As mentioned in \cref{sec:IMFsampling}, the characteristic time-scale of the SFR tracer is related to the number of star-forming regions in the emission map.
	We therefore estimate the relative change of the effective SFR tracer time-scale as the fraction of star-forming regions that do contain sufficiently massive stars to emit in the tracer of interest.
	This approach will be tested below.
	In practice, this means we need to estimate how many stars, ${N_{\min}}$, of at least some minimum mass, ${M_{\min}}$, are expected to form within a star-forming region of mass ${M_{\mathrm{r}}}$.
	We consider ${M_{\min}}$ to characterise the stellar mass at which the SFR emission becomes noticeable and not the mass contributing the most.
	The mass of the star-forming region, ${M_{\mathrm{r}}}$, can then act as a proxy for how well the IMF is sampled: smaller values of ${M_{\mathrm{r}}}$ will result in a region with an IMF that is less well sampled.

	We can calculate the probability, ${P}$, of producing a minimum number of stars ${N_{\min}}$ of at least some minimum mass ${M_{\min}}$ in a given star-forming region through a Bernoulli (i.e.\ binomial) trial.
	If the region can produce a sufficient number of stars of sufficient mass, then the region is identifiable in the SFR tracer; therefore, in our binomial trial, we define a \enquote{success} as producing a star of mass ${M}$ which satisfies the condition of ${M_{\min} \leq M \leq M_{\mathrm{r}}}$.
	The probability of success is given by ${p}$, ${N_{\star}}$ is the total number of stars within the star-forming region, and ${N}$ counts the number of \enquote{successful} stars.

	The binomial distribution gives the probability of ${k}$ successful stars:
	\begin{equation}\label{eq:binomial}
		P\!\left( N = k \right) = \frac{N_{\star}!}{k! \left(N_{\star} - k \right) !} p^{k} {\left( 1 - p \right)}^{N_{\star} - k}~\mathrm{.}
	\end{equation}
	The probability that we wish to calculate (at least ${N_{\min}}$ stars of a mass of ${M_{\min}}$ or higher) is given by
	\begin{align}\label{eq:probability}
		P\!\left( N \geq N_{\min} \right)
		& = 1 - P\!\left( N < N_{\min} \right)\\
		& = 1 - \sum_{k = 0}^{N_{\min}-1}P\!\left( N = k \right)~\mathrm{.}
	\end{align}
	The IMF, ${\mathrm{d}n/\mathrm{d}m}$, describes the distribution of mass amongst the stars within a star-forming region; this means we can use the IMF to determine the values of ${p}$ and ${N_{\star}}$ and therefore to calculate ${P\!\left( N = k \right)}$.
	In a star-forming region with a well-sampled IMF, ${p}$ is the fraction of stars that satisfy the condition ${M_{\min} \leq M \leq M_{\mathrm{r}}}$ and ${N_{\star}}$ is the total number of stars within the region:
	\begin{align}\label{eq:pn}
		&p = \nu \int_{M_{\min}}^{M_{\mathrm{r}}} \frac{\mathrm{d} n}{\mathrm{d} m} \mathrm{d} m~\mathrm{;}
		&N_{\star} = \mu \int_{0}^{M_{\mathrm{r}}} \frac{\mathrm{d} n}{\mathrm{d} m} \mathrm{d} m~\mathrm{.}
	\end{align}
	The normalisation constants ${\nu}$ and ${\mu}$ are evaluated through
	\begin{align}\label{eq:normalisation}
		&1 = \nu \int_0^{M_{\mathrm{r}}} \frac{\mathrm{d} n}{\mathrm{d} m} \mathrm{d} m~\mathrm{;}
		&M_{\mathrm{r}} = \mu \int_{0}^{M_{\mathrm{r}}} m \frac{\mathrm{d} n}{\mathrm{d} m} \mathrm{d} m~\mathrm{.}
	\end{align}
	In order to convert the probability value, ${P\!\left( N \geq N_{\min} \right)}$, into an estimate for the characteristic time-scale, ${t_{\mathrm{E}}}$, we assume a \citet{CHAB05} IMF and use the characteristic time-scales we find for a fully sampled IMF, ${t_{\mathrm{E,\,0}}}$, (see \cref{app:completeTable}) in the following equation
	\begin{equation}\label{eq:emissionTime}
		t_{\mathrm{E}} = t_{\mathrm{E, 0}} \times P\!\left( N \geq N_{\min} \right)~\mathrm{.}
	\end{equation}

	\begin{figure*}
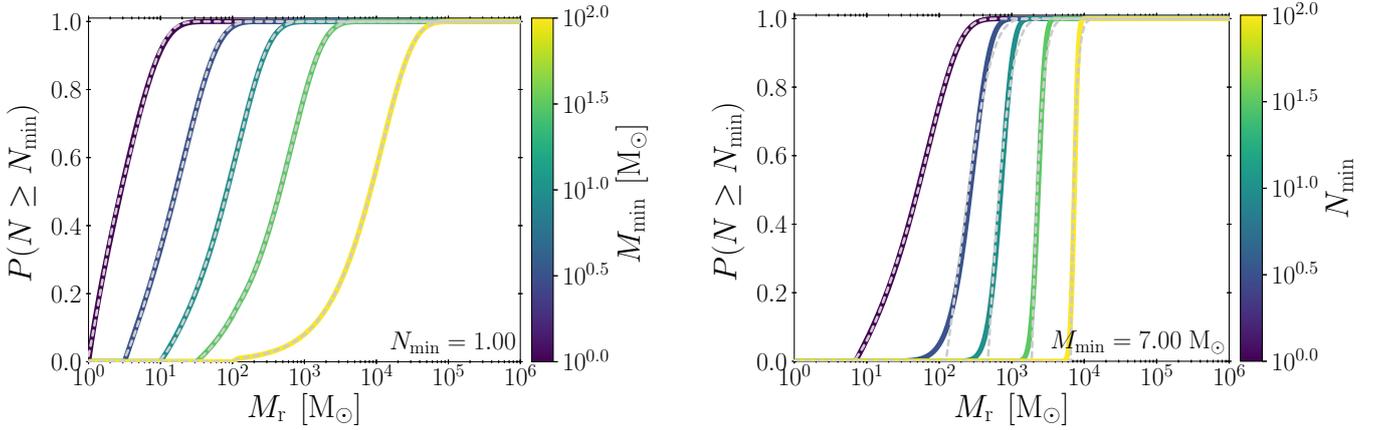

		\centering
		\includegraphics[width=\columnwidth]{\img{Nmin=1_curves}}%
		\hfill%
		\includegraphics[width=\columnwidth]{\img{Mmin=7_curves}}%
		\caption{%
			Curves showing how the probability of forming at least ${N_{\min}}$ stars of mass ${M_{\min}}$ or higher changes with star-forming region mass, ${M_{\mathrm{r}}}$.
            The grey dashed lines indicate the approximate fit to the full calculation.
            See \cref{sec:IMFsamplingTheory} for details and \cref{tab:fitParamValues} for fit parameter values.
			\textbf{Left:}~Constant ${N_{\min}}$.
			\textbf{Right:}~Constant ${M_{\min}}$.
		}\label{fig:theoModel}
	\end{figure*}

	In \cref{fig:theoModel}, we show how the form of the probability curve ${P\!\left( N \geq N_{\min} \right)}$ changes for different values of ${N_{\min}}$ and ${M_{\min}}$.
	Increasing the value of ${M_{\min}}$ increases the star-forming region mass required to reach a given probability of forming enough sufficiently massive stars (set by ${N_{\min}}$ and ${M_{\min}}$); the same effect is observed for ${N_{\min}}$ but less pronounced.
	Higher ${N_{\min}}$ also affects the probability of forming enough sufficiently massive stars by increasing the rate of change of probability with changing star-forming region mass.

	The curves in \cref{fig:theoModel} have a complex analytical form, therefore we provide a four-parameter function that approximates these curves.
	These approximations are also included in \cref{fig:theoModel} as dotted grey lines.
	The following set of equations describe the form of the approximation,
	\begin{equation}\label{eq:Mrmin}
		M_{0} = N_{\min} \times M_{\min}~\mathrm{,}
	\end{equation}
	\begin{equation}\label{eq:fm}
		\begin{split}
			f(M_{\mathrm{r}}) = 1 & + a_{1} \exp {\left(b_{1} \left[ \frac{M_{\mathrm{r}}}{M_{0}} \right] \right)} \\
			& + a_{2} \exp {\left( b_{2} {\left[ \frac{M_{\mathrm{r}}}{M_{0}} \right]}^{2} \right)}~\mathrm{,}
		\end{split}
	\end{equation}
	\begin{align}\label{eq:ApproxP}
		P\!\left( N \geq N_{\min} \right) \approx \left\{
		\begin{aligned}
			& 0 & f(M_{\mathrm{r}}) \leq 0
			\\
			& f(M_{\mathrm{r}}) & 0 < f(M_{\mathrm{r}}) < 1
			\\
			& 1 & f(M_{\mathrm{r}}) \geq 1
		\end{aligned}
		\right.~\mathrm{,}
	\end{align}
	where ${a_{i},~b_{i}}$ for ${ i=\left\lbrace 1,~2\right\rbrace }$ are four parameters that we determine through least-squares minimization. We present the parameter values for all the approximate curves displayed in \cref{fig:theoModel} in \cref{tab:fitParamValues}.
	For intermediate values of ${M_{\min}}$, these best-fitting parameters can be interpolated as a function of ${\log\left({M_{\min}}\right)}$.
	The approximate expression gives an almost identical fit in the cases where ${N_{\min} = 1}$ (see \cref{fig:theoModel}) but for higher values of ${N_{\min}}$ the approximation does not perform as well.
	Fortunately, as we show in \cref{sec:IMFsampling}, we only need to consider the case of ${N_{\min} = 1}$.

	\begin{table}
		\centering
		\caption{%
			Values for the free parameters, ${a_{i}}$ and ${b_{i}}$, in the analytical models presented in \cref{fig:theoModel} and described by \cref{eq:Mrmin,eq:fm,eq:ApproxP}
		}\label{tab:fitParamValues}
		\begin{tabular}{ll rrrr}
			\toprule
			${N_{\min}}$ & ${M_{\min}}$ & ${a_{1}}$ & ${b_{1}}$ & ${a_{2}}$ & ${b_{2}}$\tabularnewline
			\midrule
			1.000 & 1.000 & -1.001 & -0.251 & -0.384 & -0.632\tabularnewline
			1.000 & 3.162 & -1.016 & -0.137 & -0.146 & -0.338\tabularnewline
			1.000 & 10.000 & -1.021 & -0.085 & -0.071 & -0.235 \\
			1.000 & 31.623 & -1.000 & -0.048 & -0.056 & -0.233\tabularnewline
			1.000 & 100.000 & -1.000 & -0.008 & -2.393 & -5.405\tabularnewline
			\cmidrule{1-6}
			1.000 & 7.000 & -1.020 & -0.098 & -0.088 & -0.257\tabularnewline
			3.162 & 7.000 & -1.930 & -0.118 & -0.966 & -1.555\tabularnewline
			10.000 & 7.000 & -4.816 & -0.235 & -0.987 & -1.638\tabularnewline
			31.623 & 7.000 & -38.108 & -0.426 & -1.016 & -0.891\tabularnewline
			100.000 & 7.000 & -1000.000 & -0.743 & -1.016 & -0.889\tabularnewline
			\bottomrule
		\end{tabular}
	\end{table}

	We now have a description of how the characteristic time-scale of SFR tracers in a star-forming region with a stochastically sampled IMF, ${t_{\mathrm{E}}}$, is related to the characteristic time-scale determined when the IMF is well sampled, ${t_{\mathrm{E,\,0}}}$, through a probability distribution function, ${P\!\left( N \geq N_{\min} \right)}$.
	The IMF and two free parameters, ${N_{\min}}$ and ${M_{\min}}$, characterise the form of ${P\!\left( N \geq N_{\min} \right)}$.
	We note that the analytical expression for the time-scale correction factor ${P\!\left( N \geq N_{\min} \right)}$ does not carry an explicit metallicity dependence.
	We therefore apply the same theoretical framework for all metallicities, allowing us to combine the effects of both metallicity and IMF sampling on the characteristic SFR time-scale.

	\section{Error Propagation}\label{app:ErrorProp}
	In \cref{sec:IMFsamplingMethod}, we calculate the average independent star-forming region mass as
	\begin{equation}\label{eq:aisfrm}
		\overline{M}_{\mathrm{r}} = \Sigma_{\mathrm{SFR}} \times \left(t_{\mathrm{emi}} + t_{\mathrm{ref}} - t_{\mathrm{over}}\right) \times \uppi {\left(\frac{\lambda}{2}\right)}^{2}~\mathrm{.}
	\end{equation}
	This equation uses the SFR surface density, ${\Sigma_{\mathrm{SFR}}}$, and the duration of the reference map, ${t_{\mathrm{ref}}}$, along with quantities that the \textsc{Heisenberg} code measures: the typical separation length of independent star-forming regions, ${\lambda}$; the duration of the emission map, ${t_{\mathrm{emi}}}$; and the duration of the overlap between the emission and reference phases, ${t_{\mathrm{over}}}$.
	We note that \cref{eq:aisfrm} and \cref{eq:aveRegionMass} are equivalent through the definition
	\begin{equation}
		\tau \equiv t_{\mathrm{emi}} + t_{\mathrm{ref}} - t_{\mathrm{over}}~\mathrm{.}
	\end{equation}
	Here we describe how we propagate the uncertainties on these quantities into an uncertainty on the characteristic region mass ${\overline{M}_{\mathrm{r}}}$.

	To calculate the uncertainty on ${\overline{M}_{\mathrm{r}}}$ we start with the general expression: the uncertainty on a quantity ${f}$, ${\sigma_{f}}$, which is a function of ${N}$ variables i.e.\ ${f\left(x_{1}, \ldots, x_{N} \right)}$ is given by \citep{HUGH10}
	\begin{align}\label{eq:errorGen}
		\sigma_{f}^{2} =
		\sum_{i = 1}^{N} \sum_{j = 1}^{N} \frac{\partial f}{\partial x_{i}} \frac{\partial f}{\partial x_{j}} \rho_{ij} \sigma_{i} \sigma_{j}~\mathrm{,}
	\end{align}
	where ${\sigma_{i}}$ represents the uncertainty on variable ${x_{i}}$ and ${\rho_{ij}}$ represents correlation coefficients between variable ${x_{i}}$ and ${x_{j}}$ (where ${\rho_{ii} = 1}$ and ${\rho_{ij} = \rho_{ji}}$).
	In order to simplify our expressions and to use the same notation as in \cref{eq:errorGen}, we define the following
	\begin{align}
		\kappa &\equiv \Sigma_{\mathrm{SFR}} \, \frac{\uppi}{4}~\mathrm{,} \\
		x_{1} &\equiv \lambda~\mathrm{,} \\
		x_{2} &\equiv t_{\mathrm{emi}}~\mathrm{,} \\
		x_{3} &\equiv t_{\mathrm{over}}
	\end{align}
	and \cref{eq:aisfrm} becomes
	\begin{align}
		\overline{M}_{\mathrm{r}} & = \kappa \tau {x_{1}}^{2} \\
		& = \kappa \left(x_{2} + t_{\mathrm{ref}} - x_{3}\right) {x_{1}}^{2}~\mathrm{.}
	\end{align}
	We note that ${\Sigma_{\mathrm{SFR}}}$ and ${t_{\mathrm{ref}}}$ are considered to be without error and do not need to be included as variables.
	The derivatives we need in order to calculate ${\sigma_{\overline{M}_{\mathrm{r}}}}$ are
	\begin{align}\label{eq:deriv}
		\frac{\partial}{\partial x_{1}}\,\overline{M}_{\mathrm{r}} & = 2 \kappa \tau x_{1}~\mathrm{,} \\
		\frac{\partial}{\partial x_{2}}\,\overline{M}_{\mathrm{r}} & = \kappa {x_{1}}^{2}~\mathrm{,} \\
		\frac{\partial}{\partial x_{3}}\,\overline{M}_{\mathrm{r}} & = - \kappa {x_{1}}^{2}~\mathrm{.}
	\end{align}
	Combining \cref{eq:deriv} with \cref{eq:errorGen} we find the expression for the uncertainty on ${\overline{M}_{\mathrm{r}}}$, ${\sigma_{\overline{M}_{\mathrm{r}}}}$:
	\begin{equation}
		\begin{split}
			{\left[{\frac{\sigma_{\overline{M}_{\mathrm{r}}}}{\overline{M}_{\mathrm{r}}}}\right]}^{2} & =  \frac{4 {\sigma_{1}}^{2}}{{x_{1}}^{2}} \\
			& + \frac{\left(
				{\sigma_{2}}^{2} + {\sigma_{3}}^{2} - 2 \rho_{23} \sigma_{2} \sigma_{3}
				\right)}{\tau^{2}} \\
			& + \frac{4 \left(\rho_{12}\sigma_{1}\sigma_{2} - \rho_{13}\sigma_{1}\sigma_{3}\right)}{\tau x_{1}}~\mathrm{.}
		\end{split}
	\end{equation}
	With this expression, we can take into account the associated uncertainty on the value of ${\overline{M}_{\mathrm{r}}}$ as part of our error analysis and $\chi^{2}$ calculations when investigating the effects of incomplete IMF sampling on the characteristic time-scales of SFR tracers.

	\bsp	
	\label{lastpage}
\end{document}